\documentclass[a4paper,fleqn]{cas-sc}

\usepackage[numbers]{natbib}

\usepackage{svg}

\usepackage{caption}
\usepackage{balance}
\usepackage{amsmath}
\usepackage{graphicx} 
\usepackage{fancyhdr}
\usepackage[english]{babel}

\usepackage{hyperref}

\usepackage{booktabs}

\usepackage{booktabs}

\usepackage[draft]{minted}
\newenvironment{code}{\captionsetup{type=listing}}{}
\usepackage{caption}
\usepackage{newfloat}

\definecolor{light-gray}{gray}{0.95}
\newcommand{\codehighlight}[1]{\colorbox{light-gray}{\small\texttt{#1}}}
\newtheorem{definition}{Definition}
\usepackage{csquotes}
\usepackage{subfigure}
\usepackage{svg}
\usepackage{eurosym}
\usepackage{float}

\def\tsc#1{\csdef{#1}{\textsc{\lowercase{#1}}\xspace}}
\tsc{WGM}
\tsc{QE}

\begin{document}
\let\WriteBookmarks\relax
\def\floatpagepagefraction{1}
\def\textpagefraction{.001}

\shorttitle{\textit{trasgoDP}: An Open Source Framework for Releasing Noised Tabular Microdata under Local Differential Privacy}    

\shortauthors{J. S\'ainz-Pardo D\'iaz \& \'A. L\'opez Garc\'ia}  

\title[mode = title]{\textit{trasgoDP}: An Open Source Framework for Releasing Noised Tabular Microdata under Local Differential Privacy}

\author[1]{Judith {S\'ainz-Pardo D\'iaz}}[orcid=0000-0002-8387-578X] \ead{sainzpardo@ifca.es} \cormark[1]\credit{Conceptualization, 
Formal analysis,
Investigation, 
Methodology, 
Software, 
Validation, 
Visualization, 
Writing - original draft}

  \author[1]{{\'A}lvaro {L\'opez Garc\'ia}}[orcid=0000-0002-0013-4602] \ead{aloga@ifca.es}\credit{Funding acquisition,
Resources,
Supervision,
Validation,
Project administration,
Writing - review \& editing}

\affiliation[1]{organization={Instituto de F\'isica de Cantabria (IFCA), CSIC-UC},
  addressline={Avda. los Castros s/n},
  city={Santander},
  postcode={39005},
  country={Spain}
}


\begin{abstract}
\textit{trasgoDP} is a modular, open-source, and easy-to-use Python framework for releasing tabular microdata under $\epsilon$-local differential privacy guarantees, as well as location data under geo-indistinguishability assumptions, designed to be installed and integrated within standard data science workflows. The software enables systematic exploration of privacy-utility trade-offs across multiple mechanisms, data types, and $\epsilon$ values. While differential privacy has been extensively studied for aggregate data, its application to row-wise microdata release remains underexploited in terms of reusable software tools, a gap that is even more pronounced in the case of metric privacy and location-based data. \textit{trasgoDP} implements local-DP mechanisms for numerical and categorical attributes (Laplace, Gaussian, Exponential, and Randomized Response), a geo-indistinguishability mechanism for location data, and a set of utility metrics, including a novel correlation-loss measure, to quantify information loss as a function of the allocated privacy budget. The objective of this work is to provide the research community with a reproducible, open-source baseline for evaluating tabular and location-based data publication methodologies under formal local differential privacy guarantees.
\end{abstract}


\maketitle

\section*{INTRODUCTION}

The publication of open data is a fundamental pillar for building open science\cite{bertram2023open, ramachandran2021open}. More specifically, the three main pillars that compose open science are open data, open source, and open access.  
As data availability continues to grow, increasingly larger datasets are being released daily, allowing the development of a wide range of applications, including predictive models based on machine learning (ML) techniques. However, in certain domains, the publication of data is a complex process, especially when sensitive data or data that could identify individuals are involved and carry the risk of individual re-identification. In this sense, this challenge becomes even more complex in the context of microdata\cite{ciriani2007microdata} releasing, where tabular data\cite{domingo2022anonymization} presents an additional complexity: the anonymization guarantees applied to an early publication may be compromised by later updates that introduce new records.

On the one hand, when we refer to microdata publication, we are dealing with raw data containing individual records on a group of studied individuals. This could be a hospital patient record, a municipal census, or data on public employees. On the other hand, when we talk about aggregated data, we are referring to statistics compiled from that microdata. In some cases, publishing aggregate data can be useful, but in others cases it is essential to have the complete dataset, in order to perform advanced tasks such as classification, clustering, or predictive modeling. In the same line, location data represents a particularly sensitive case of microdata. Each record typically contains precise latitude and longitude coordinates tied to an individual or event, and simple aggregation is often not sufficient to protect privacy, since even a small number of location points can reveal sensitive patterns such as home addresses, workplaces, etc. 

Usually, when releasing tabular microdata, the applied anonymization techniques focus on three types of attributes: identifiers (ID), which must be removed prior to publication; quasi-identifiers (QIs), which are anonymized through generalization; and sensitive attributes (SAs) which are protected by generalizing the QIs to avoid their association with an identified or identifiable individual. However, we can consider the case of releasing hospital patient records, where QIs such as age, gender, and demographic data are used as QIs, and the reason for admission is used as SA. Having this data is relevant for building a wide range concerning predictive and personalized health. However, what happens if we anonymize the microdata, publish it, and a week later a new (anonymized) release is published including new records? In this case, it is possible that an attacker could extract relevant information about the new records by analyzing that difference. 

In this sense, the idea of applying differential privacy (DP)\cite{dwork2014algorithmic} arises naturally, as it provides a formal definition for the privacy ensured with the guarantee that whatever additional information the attacker has, the level of privacy guaranteed by the privacy budget ($\epsilon$) cannot be broken. 

DP is usually applied in its global form, i.e., we could think of applying DP to build aggregated data (when calculating statistics). This is evident if we consider the amount of noise we add in one case or the other (just taking into account the central limit theorem). Thus, although local differential privacy\cite{xiong2020comprehensive} has been studied theoretically, it is not widely implemented in the case of microdata publication, especially due to the high level of noise required to disturb the data. In local-DP, it is essential to strike a balance between the level of privacy applied and the amount of information we need to maintain so that the data remains statistically significant. 

In this line, with the aim of providing the community with tools that allow for the evaluation of such trade-offs, and whether the use of local-DP notions can be useful for such publication (taking into account the composition when applied to more than one column), we observe a lack of open tools for this purpose, especially if we think of those written in Python that allow us to easily incorporate it into data science, processing, visualization, and analysis flows. This gap is even more pronounced when working with location-based data and metric-privacy or geo-indistinguishability notions. This led us to the implementation of the open source Python library \texttt{trasgoDP}, which implements local-DP mechanisms, a geo-indistinguishability model for metric privacy, and utility metrics. It aims to complement \texttt{pyCANON}\cite{sainzpardo2022pycanon} and \texttt{anjana}\cite{sainzpardo2024anjana}, two Python libraries for anonymizing and checking the level of anonymity in tabular datasets, respectively. It forms the third part of a toolkit that seeks to enable the secure publication of microdata, because to promote open science, we need open data but also open source tools. 

\subsection*{Related work}

When we refer to data privacy, we are referring to the protection of information from unauthorized use and access. In the current context, this is particularly relevant due to the massive generation and storage of data, which increases the risk of re-identification of individuals, data leaks, linking of data from different databases, etc\cite{domingo2016database}. 

As mentioned previously, anonymization methods are widely used for this purpose, particularly in accordance with the GDPR principle of \emph{data minimization}, which states that anonymized data is no longer considered personal data related to an identified or identifiable individual. In this regard, there are different software products that enable the effective anonymization of tabular datasets, such as \texttt{ARX}\cite{Prasser_Flexible_data_anonymization_2020} or \texttt{anjana}; for protecting statistical tables, such as \texttt{$\tau$-argus} or for creating microdata files, such as \texttt{$\mu$-argus}. 

However, in some cases, the goal is not to anonymize the data, but rather to apply other measures that protect it without resorting to the use of hierarchies, generalization or suppression methods. In this regard, we are interested in generating perturbed data (following the idea of synthetic data generation but starting from the raw data) using techniques based on differential privacy, thereby creating privatized versions of the original records.

When it comes to software related to the implementation of mechanisms that ensure differential privacy, we must first review those that focus on global-DP, as these mechanisms are widely adopted due to their utility and numerous distributions are available. Specifically, we can highlight the following libraries: \texttt{OpenDP}\cite{Shoemate_OpenDP_Library}, which is a modular Python library that implements a suite of statistical algorithms that satisfy the definition of differential privacy. \texttt{pyDP}\cite{pydp} is a Python wrapper for Google's Differential Privacy created by OpenMined. It provides differentially private algorithms, including statistics such as the mean, median, percentiles, etc., bounded by DP. The methods implemented use the Laplacian mechanism. In addition, \texttt{diffprivlib}\cite{diffprivlib} is the IBM differential privacy library written in Python, which implements the Laplace, Gaussian, Exponential, and randomized response methods for global DP, creating sanitized histograms, and for training ML models, including supervised and unsupervised learning. 

However, when we look for tools that allow us to apply DP mechanisms to raw data, the number of open-source solutions available is much smaller. Specifically, we can highlight three Python libraries: \texttt{Multi-Freq-LDPy}\cite{arcolezi2022multi}, which allows to perform multiple frequency estimation tasks under LDP guarantees; \texttt{LDP-Toolbox}\cite{10.1145/3719027.3760706}, that explores utility and attacks trade-offs in local-DP, and \texttt{pure-LDP}\cite{cormode2021frequency}, which provides simple implementations of state-of-the-art LDP frequency estimation algorithms. 

More specifically, regarding location-based data, the implications of introducing metrics and distance notions to the differential privacy paradigm was proposed in 2013 by Chatzikokolakis et al. in \cite{chatzikokolakis2013broadening}. Then, the notion of geo-indistinguishability was introduced in by Andr\'es et al with the idea of adding random noise to the user's location from a planar Laplace distribution \cite{andres2013geo}. In addition, this research line has been extended with new algorithms that aim to solve the problems regarding the protection of isolated locations by the Laplace mechanism \cite{biswas2024privic}. However, to the best of our knowledge, no actively maintained, general-purpose open-source Python library currently provides an accessible implementation of such mechanisms integrated within a broader local-DP toolkit focused on tabular microdata releasing.

In light of this, it was decided to implement \texttt{trasgoDP}, which implements algorithms for LDP applied directly to categorical and numerical data (taking into account the limitations in terms of the amount of noise to be added), a mechanism for geo-indistinguishability and thus metric privacy, as well as functions that allow us to measure and compare the utility of DP-protected data versus raw data, and to quantify the divergence between them. The goal is to build a comprehensive toolkit for data privatization alongside related tools such as anjana, which anonymizes tabular data, and \texttt{pyCANON}, which determines the level of anonymity of a dataset. Thus, \texttt{trasgoDP} forms the third component of this modular toolkit written in Python, which can be easily integrated into a data science pipeline.

\subsection*{Data releasing}
\begin{figure*}
\centering
\includegraphics[width=0.8\linewidth]{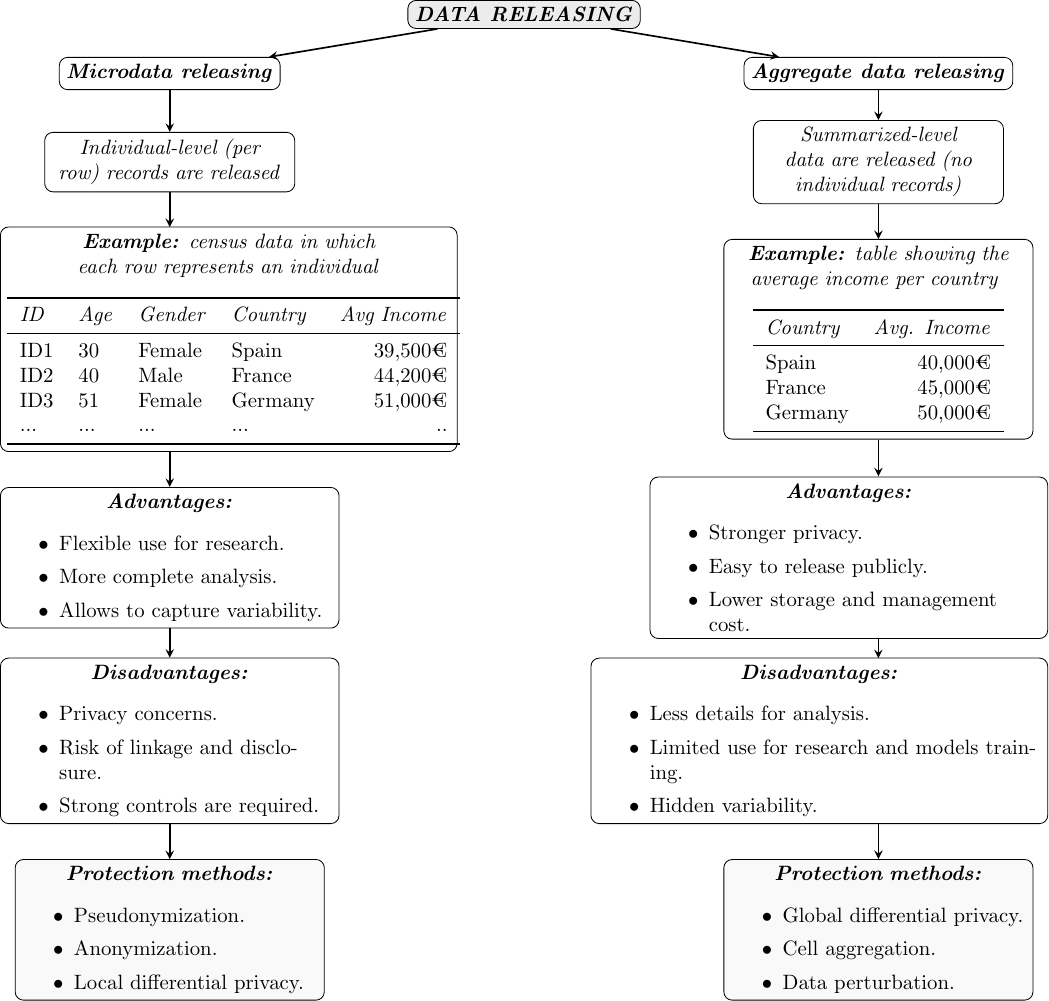}
\caption{Difference between microdata releasing and aggregate data publishing.}
\label{fig:diagram_microdata_aggregate}

\end{figure*}

We have already discussed the difference between microdata and aggregated data, but it is important to elaborate on this point and compare both aspects to highlight why we are sometimes interested in real microdata releasing and thus why it is important when working on differential privacy to have tools not only for global-DP, but also for local-DP. Specifically, the differences between these two aspects are summarized in the diagram in Figure~\ref{fig:diagram_microdata_aggregate}.

\begin{figure}
    \centering
    \includegraphics[width=0.7\linewidth]{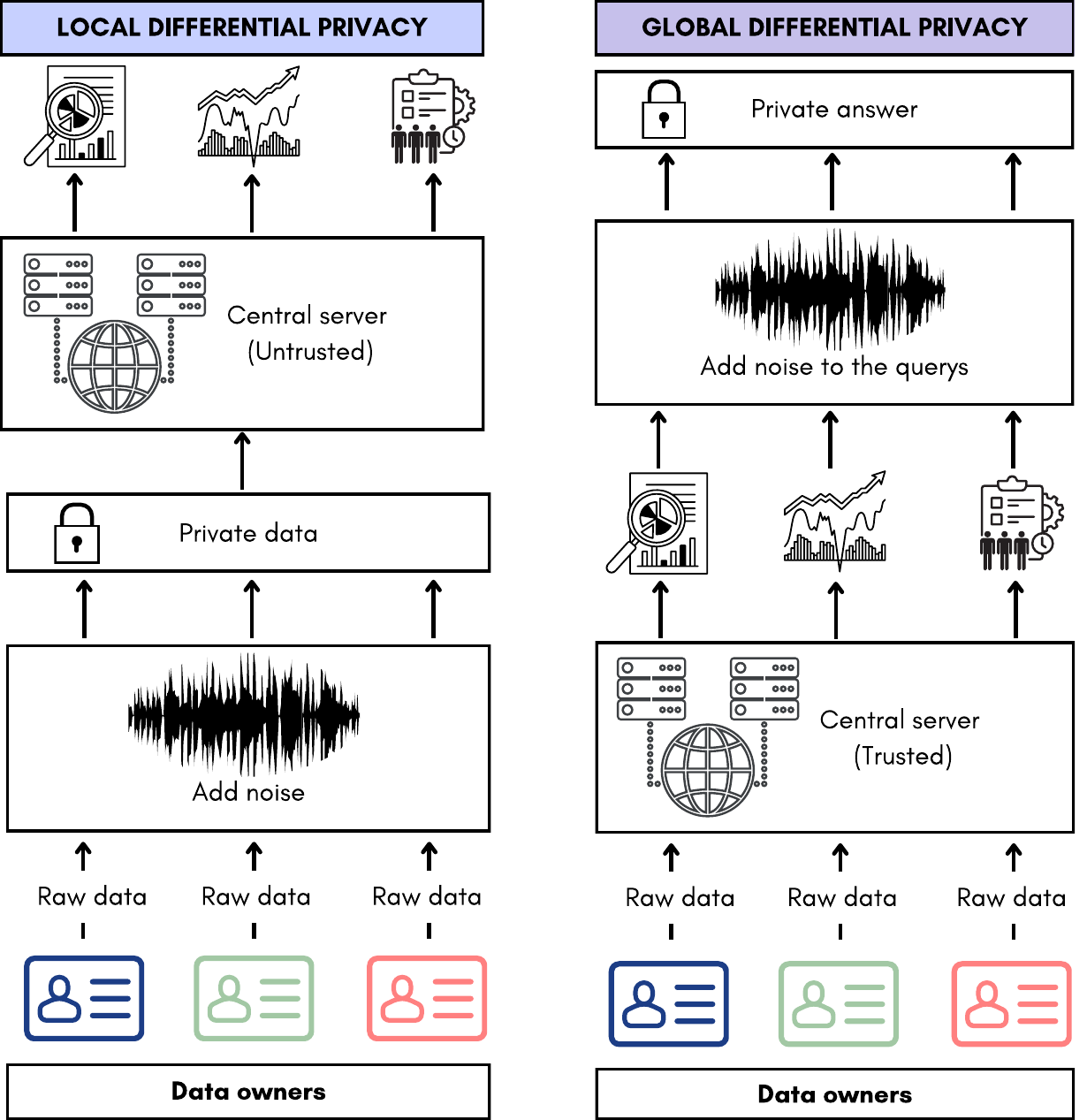}
    \caption{Local-DP vs global-DP: schematic comparison.}
    \label{fig:local_global_dp}
\end{figure}

In both cases, one method that can be used for data protection is differential privacy. As already mention, we can think on both local and global versions, depending on which step of the data processing pipeline we apply it. The workflows for these two approaches are compared in Figure \ref{fig:local_global_dp}. 

Then, let's start giving the formal definition for this theoretical guarantee. Here it is important to take into account that differential privacy is not a method or algorithm itself, but a mathematical definition.

\subsection*{Differential privacy}

As in this work we are focusing in local-DP instead of in global-DP, the following definitions will be given for the discrete case. 

\begin{definition}{\textbf{Local} $\boldsymbol{\epsilon}$\textbf{-differential privacy.}}
A randomized algorithm $\mathcal{M}$, with domain $\mathcal{D}$ and range $\mathcal{R}$, satisfies \textbf{local} $\boldsymbol{\epsilon}$\textbf{-differential privacy} if for any inputs $y, y' \in \mathcal{D}$ and for any possible output $r\in \mathcal{R}$ it is satisfied that:

$$
\mathbb{P}[\mathcal{M}(y)=r] \leq e^{\epsilon}\mathbb{P}[\mathcal{M}(y')=r], \mbox{ with } \epsilon\geq 0.
$$
\end{definition}

\begin{definition}{\textbf{Local} $\boldsymbol{(\epsilon,\delta)}$\textbf{-differential privacy.}}
A randomized algorithm $\mathcal{M}$, with domain $\mathcal{D}$ and range $\mathcal{R}$, satisfies \textbf{local} $\boldsymbol{(\epsilon,\delta)}$\textbf{-differential privacy} if for any inputs $y, y' \in \mathcal{D}$ and for any possible output $r\in \mathcal{R}$ it is satisfied that:

$$
\mathbb{P}[\mathcal{M}(y)=r] \leq e^{\epsilon}\mathbb{P}[\mathcal{M}(y')=r]+\delta, \mbox{ with } \epsilon\geq 0 \mbox{ and } \delta\in[0,1].
$$
\end{definition}

Concerning ensuring DP guarantees, different mechanisms can be considered. Specifically, let's review the first ones implemented in \texttt{trasgoDP} in version v2.0.1 for both numerical and categorical attributes. 

\paragraph*{Numerical attributes}
The following mechanism adopted for numerical attributes are usually applied in global-DP scenarios, but we can easily adopt them for the local case. Specifically, in the following definitions we need to define a proper notion of the sensitivity and we can consider the function $f(\cdot$) as the identity applied on each record. Then, is a widely known result that for ensuring $\epsilon$-global-DP, it is proven that the following mechanism (Laplace mechanism) can be adopted:

        \begin{definition}{\textbf{Laplace mechanism.}}\label{def:laplace}
        Given any function $f:\mathcal{D} \longrightarrow \mathbb{R}^{k}$, we define the \textbf{Laplace mechanism} as:
        $$
        \displaystyle
        \mathcal{M}_{L}(x, f(\cdot), \epsilon):= f(x) + (Y_{1}, \hdots, Y_{k}),
        $$
        where $Y_{i},  \forall i \in \{1, \hdots, k\}$ are independent and identically distributed (i.i.d.) variables from the Laplace distribution with location 0 and scale $\Delta_{1} (f)/\epsilon$ (equivalently $Laplace(0, \Delta_{1} (f)/\epsilon)$), being $\Delta_{1}(f)$ the $l_{1}$-sensitivity.
        \end{definition}

 Concerning $(\epsilon,\delta)$-global-DP guarantees the Gaussian mechanism can be applied:
        \begin{definition}{\textbf{Gaussian mechanism.}}\label{def:gaussian}
        Given any function $f:\mathcal{D} \longrightarrow \mathbb{R}^{k}$, we can define the \textbf{Gaussian mechanism} as:
        $$
        \displaystyle
        \mathcal{M}_{G}(x, f(\cdot), \epsilon, \delta):= f(x) + (Y_{1}, \hdots, Y_{k}),
        $$
        where $Y_{i},  \forall i \in \{1, \hdots, k\}$ are independent and identically distributed (i.i.d.) variables from the Gaussian distribution $N(0, \sigma^2)$, with $\sigma = \frac{\Delta_{2}(f) \sqrt{2\log(1.25/\delta)}}{\epsilon}$, being $\Delta_{2}(f)$ the $l_{2}$-sensitivity.
        \end{definition}     

Note that for moving to the local-DP case, and assuming that we are dealing with bounded numerical data (e.g. $x \in  [a,b]$), then $\Delta_{1}=\Delta_{2}=b-a$.

In view of the above, it is important to note that the magnitude of the noise required differs substantially between the central (or global) model and the local model of differential privacy. In the case of global differential privacy (central model), when aggregated statistics such as the mean are privatized, sensitivity scales as $(b-a)/n$. Consequently, the noise added by Laplace or Gaussian mechanisms decreases proportionally to $1/n$. This implies that as the sample size increases, the necessary perturbation decreases and statistical utility improves significantly. On the contrary, in the local differential privacy model, each individual perturbates their own data before sharing it, and in this case the sensitivity is $b-a$, independent of the sample size. Therefore, the magnitude of the noise does not depend on n, which leads to a significantly higher variance compared to the global approach. This difference is the fundamental mathematical reason why mechanisms under LDP are considerably noisier than under the global assumption.

In this line, we also have to note that, as the Gaussian mechanism provides ($\epsilon$, $\delta$)-guarantees, in the local-DP approach we have to carefully select the value of the parameter $\delta$. Specifically, an initial approach should be to fix $\delta << 1/n$, with $n$ the number of records of the dataset.

However, as already stated, the local approach may be particularly relevant in scenarios where there is no central trusted entity, when users want to maintain direct control over their data, or in distributed systems for mass information collection. Likewise, LDP is particularly suitable for the construction of private histograms or frequency estimation, specifically in cases in which the large number of participants partially compensates for the increase in variance induced by individual disturbance.
    
\paragraph*{Categorical attributes} 
For categorical attributes, privacy preservation under the local model can be achieved through mechanisms specifically designed for discrete domains. In this case, we can apply the Exponential mechanism or the Randomized Response one. Again, as in the numerical setting, we need to take into account the trade-off between privacy and utility, concerning the amount of noise added compared to the global approach, as in this setting each records is perturbed independently. 

Note that for the Randomized Response mechanism we can define two strategies depending weather or not we are dealing with binary attributes. For the non binary case we will call this method $k$-ary Randomized Response. These three mechanisms are proven to provide $\epsilon$-DP guarantees.

\begin{definition}{\textbf{Exponential mechanism.}}
    Be $D$ the set of inputs, $\mathcal{R}$ the set of outputs, $r\in \mathcal{R}$ and $\Delta(g)$ the sensitivity of the utility or score function $g$. The \textbf{exponential mechanism} outputs $r$ with probability $\mathbb{P}[r]$ defined as follows:

$$
\mathbb{P}[r]=\frac{exp\left(\frac{\epsilon\cdot g(D,r)}{2\Delta(g)}\right)}{\sum_{r'\in\mathcal{R}}exp\left(\frac{\epsilon\cdot g(D,r')}{2\Delta(g)}\right)}
$$
\end{definition}

\begin{definition}{\textbf{Randomized Response (RR) mechanism for binary attributes.}} Be $\mathcal{D}=\{y, \tilde{y}\}$) the data domain. Given the privacy budget $\epsilon$, be $p(\epsilon)=\frac{e^{\epsilon}}{e^{\epsilon}+1}$, the \textbf{randomized response algorithm for binary attributes} for the value $y$ given $\epsilon$ returns $\overline{y}$ as follows (note that $\tilde{y}=\neg y$): 

    $$    \overline{y}=\left\{ \begin{array}{l}
             y \hspace{1cm} \mbox{with probability } p(\epsilon) \\ 
             \neg y \hspace{0.8cm} \mbox{with probability } 1-p(\epsilon) \\ 
             \end{array}
    \right..
$$

Note that if $\epsilon=0$, $p(\epsilon)=0.5$, and $lim_{\epsilon \rightarrow +\infty}p(\epsilon)=1$.

\end{definition}

\begin{definition}{\textbf{k-ary Randomized Response (RR) mechanism.}}
    Be $k$ the number of different values in the data domain ($\mathcal{D}=\{y_{1}, \hdots, y_{k}\}$). The \textbf{k-ary randomized response algorithm} for the value $y$ given $\epsilon$ is the true value $y$ if $b=0$ and otherwise it is sampled from $\mathcal{D}$ following an uniform distribution. In this approach we get $b \sim Ber(k/(e^{\epsilon}+k-1))$, with \textit{Ber} the Bernoulli distribution.
\end{definition}

The three mechanisms described above are proposed as perturbation strategies applied independently to individual records of a categorical column in a dataframe or categorical array. Within the \texttt{trasgoDP} library, they are implemented as local privatization operators, enabling the transformation of categorical attributes into synthetic or noised representations while maintaining formal $\epsilon$-DP guarantees. 

Finally, it is important to note that when LDP mechanisms are applied to multiple columns of a dataset, the overall privacy guarantee degrades according to the sequential composition theorem. This means that if we apply local-DP with $\epsilon_{i}$ $\forall i \in\{1,\hdots,n_{col}\}$ (with $n_{col}$ the number of columns in the database to be perturbed), the combined release satisfies $\sum_{i=1}^{n_{col}}\epsilon_{i}$. This implies that practitioners must treat the privacy budget as a global resource to be distributed across columns, rather than as a per-column parameter. More details concerning composition theorems in DP can be found in \cite{8049725}.

\subsection*{Metric privacy} 
When dealing with location-based data, we can naturally think of the notion of \textit{metric differential privacy} (also known as \textit{d-privacy}), and in the following referred to simply as \textit{metric privacy}. The main objective of metric privacy is to take into account the geographical distance in the development of location privacy solutions and geo-indistinguishability, in order to protect an individual's location in location-based services. It is formally defined as follows:

\begin{definition}{$\boldsymbol{\epsilon}$-\textbf{metric-privacy.}}
A randomized algorithm $\mathcal{M}$, with domain $\mathcal{D}$ and range 
$\mathcal{R}$, with $\mathcal{D}$ provided with a metric 
$d:\mathcal{D}^{2} \longrightarrow \mathbb{R}_{\geq 0}$, satisfies 
$\boldsymbol{\epsilon}$\textbf{-metric-privacy}
(see \cite{biswas2024privic}) if for any inputs $y, y' \in \mathcal{D}$ and 
for any possible output $r\in \mathcal{R}$ it is satisfied that:
$$
\mathbb{P}[\mathcal{M}(y)=r] \leq e^{\epsilon \cdot d(y,y')}\mathbb{P}[\mathcal{M}(y')=r],
$$
with $\epsilon\geq 0$. Note that we assume that $\mathcal{D}$ is provided 
with a metric space.
\end{definition}

From the previous definition we can note that the inputs that are closer in relation to the given metric $d$ will be more indistinguishability to an attacker, while the ones that are more distant will be more easily discernible. Then, we can get the following definition for the concept of geo-indistinguishability as presented in \cite{andres2013geo}.

\begin{definition}{\textbf{Geo-indistinguishability.}}
    A mechanism provides guarantees of \textbf{geo-indistinguishability} if 
    and only if for any radius $x > 0$ we can ensure \textit{$\epsilon$-metric-privacy} 
    within the radius $x$.
\end{definition}

In \texttt{trasgoDP}, geo-indistinguishability is achieved by perturbing a true location $(lat, lon)$ through the addition of a two-dimensional noise vector drawn in polar coordinates. Concretely, an angle $\theta \sim \text{Unif}[0, 2\pi)$ is drawn together with a radius $r$ 
sampled from a Gamma distribution with shape $2$ and scale $1/\epsilon$, according to \cite{andres2013geo}. Then, the mechanism implemented is defined as follows:

\begin{definition}{\textbf{Geo-indistinguishability mechanism}}\label{def:geoindis_mechanism}
Given a location $(lat, lon) \in [-90,90]\times[-180,180]$, a privacy budget $\epsilon>0$ and the Earth's radius $R_{\oplus}$ (in meters), the mechanism draws $\theta \sim \text{Unif}[0,2\pi)$ and 
$r \sim \text{Gamma}(2, 1/\epsilon)$, and outputs the perturbed location $(lat^{*}, lon^{*})$ as:
$$
lat^{*} = lat + \frac{r\cos\theta}{R_{\oplus}}\cdot\frac{180}{\pi}, 
$$
$$
lon^{*} = lon + \frac{r\sin\theta}{R_{\oplus}\cos(lat \cdot \pi/180)}\cdot\frac{180}{\pi},
$$
returning the triplet $(lat^{*}, lon^{*}, r)$.
\end{definition}

Note that according with the previous definition, the amount of noise added does not depend on the sample size $n$, but here it is further determined by the geometry of the domain.

\section*{RESULTS}

\subsection*{Local differential privacy mechanisms: use examples}
As already stated, in the initial implementation of \texttt{trasgoDP}, three mechanisms for categorical attributes and two for numerical ones are implemented. In Table~\ref{tab:functions_trasgodp}, these five methods are shown together with the name of the corresponding function in the library and the input expected for applying them. Note that in the following we show the function for the case of working directly with pandas dataframe. However, all these mechanisms have their corresponding functions for the cases in which numpy arrays or lists of values are introduced, as will be explained in the following. Note that for numerical attributes, the lower and upper bounds are parameters that must be fixed using public domain knowledge (e.g. a plausible age range) established independently of the dataset being sanitized, the same as the positive label for binary attributes. \texttt{trasgoDP} requires these bounds to be provided explicitly by the user, and does not infer them automatically from the data, as doing so would make the sensitivity data-dependent and would invalidate the formal privacy guarantees. Finally, the default $\delta$ parameter for the Gaussian mechanism, which is customizable, has been set to 1e-3.
\begin{center}
    \resizebox{\linewidth}{!}{
    \begin{tabular}{rl}
    \toprule
    \textbf{Method} & \textbf{Function and input}\\
    \midrule
    \textit{Laplace mechanism} &  \textit{dp\_clip\_laplace(df, column, epsilon, lower\_bound, upper\_bound, new\_column)}\\
    \textit{Gaussian mechanism} &  \textit{dp\_clip\_gaussian(df, column, epsilon, lower\_bound, upper\_bound, delta, new\_column)}\\
    \textit{Exponential mechanism} &  \textit{dp\_exponential(df, column, epsilon, new\_column)}\\
    \textit{RR mechanism (binary)} & \textit{dp\_randomized\_response\_binary(df, column, epsilon, new\_column, positive\_label})\\
    \textit{RR k-ary mechanism} & \textit{dp\_randomized\_response\_kary(data, epsilon)}\\
    \bottomrule
    \end{tabular}}
    \captionof{table}{Function which implement each LDP mechanism for the case of pandas dataframes.}
    \label{tab:functions_trasgodp}
\end{center}

Following the examples conducted for the cases of both \textit{pyCANON} and \textit{anjana}, in the Example Code~\ref{code:example_adult} we present an example of application of the five mechanisms to different columns of a pandas dataframe (one mechanism each time). This example is conducted for the classic \textit{adult dataset}\cite{adult_dataset}, which is an extraction of the 1994 Census database composed of 32,561 rows in the train set (used in this work). The data has been previously processed and it is available in the examples folder of the \texttt{trasgoDP} library. Note that, in the following example, in each case we create a new dataframe with the resulting column obtained when applying DP.

\begin{code}
\begin{minted}[fontsize=\small]{python}
import pandas as pd
from trasgodp.numerical import dp_clip_laplace, dp_clip_gaussian
from trasgodp.categorical import (
    dp_exponential,
    dp_randomized_response_kary,
    dp_randomized_response_binary,
)

# Read the data (already processed):
data = pd.read_csv("adult_processed.csv")

epsilon = 1
# Apply DP for the attribute age with the Laplace mechanism:
lower_bound = 16
upper_bound = 100
df_lap = dp_clip_laplace(data, "age", epsilon, lower_bound, upper_bound, new_column=True)

# Apply DP for the attribute age with the Gaussian mechanism:
df_gauss = dp_clip_gaussian(data, "age", epsilon, lower_bound, upper_bound, new_column=True)

# Apply DP for the attribute workclass with the Exponential mechanism:
df_exp = dp_exponential(data, "workclass", epsilon, new_column=True)

# Apply DP for the attribute workclass with the k-ary Randomized Response mechanism:
df_kary = dp_randomized_response_kary(data, "native-country", epsilon, new_column=True)

# Apply DP for the attribute sex with the Randomized Response mechanism (binary):
df_bin = dp_randomized_response_binary(data, "sex", epsilon, new_column=True)
\end{minted}
\caption{Example: applying the five implemented mechanisms to different columns in the case of the \textit{adult dataset}.}
\label{code:example_adult}
\end{code}
\vspace{0.3cm}

In case the functions presented in Table~\ref{tab:functions_trasgodp} should be applied directly to a list or a numpy array rather than to a pandas dataframe, Table~\ref{tab:functions_trasgodp_array} shows the functions implemented for that purpose. All of them return a numpy array containing the transformation of the original list by adding DP. 

\begin{center}
    \resizebox{\linewidth}{!}{
    \begin{tabular}{rl}
    \toprule
    \textbf{Method} & \textbf{Function and input}\\
    \midrule
    \textit{Laplace mechanism} &  \textit{dp\_laplace\_array(data, epsilon, lower\_bound, upper\_bound)}\\
    \textit{Gaussian mechanism} &  \textit{dp\_clip\_gaussian(data, epsilon, lower\_bound, upper\_bound, delta)}\\
    \textit{Exponential mechanism} &  \textit{dp\_exponential\_array(data, epsilon)}\\
    \textit{Randomized response mechanism (binary)} & \textit{dp\_randomized\_response\_binary\_array(data, epsilon, positive\_label})\\
    \textit{Randomized response k-ary mechanism} & \textit{dp\_randomized\_response\_kary\_array(data, epsilon)}\\
    \bottomrule
    \end{tabular}}
    \captionof{table}{Function which implement each LDP mechanism for the case of a list or numpy array.}
    \label{tab:functions_trasgodp_array}
\end{center}

\subsection*{Metric privacy via geo-indistinguishability}
As already introduced, \texttt{trasgoDP} implements the geo-indistinguishability mechanism presented in Definition~\ref{def:geoindis_mechanism} to sanitize pairs of latitude and longitude. Table~\ref{tab:functions_geoindis} shows the function implemented in \texttt{trasgoDP} in this regard and for plotting the results, together with its expected input for the case of a pandas dataframe. Note that unlike the mechanisms for numerical and categorical data described above, this 
mechanism does not operate on a single column but jointly perturbs a pair of coordinates

\begin{center}
    \centering
    \resizebox{\linewidth}{!}{
    \begin{tabular}{rl}
    \toprule
    \textbf{Method} & \textbf{Function and input}\\
    \midrule
    \textit{Geo-indistinguishability mechanism} &  \textit{metric\_privacy(df, column\_lat, column\_lon, epsilon, new\_cols, earth\_radius\_m, seed)}\\
    \textit{Plot the resulting map} &
    \textit{plot\_metric\_dp\_map(df\_dp, column\_lat, column\_lon, save\_file)}\\
    \bottomrule
    \end{tabular}}
    \captionof{table}{Metric privacy related functions: applying the geo-indistinguishability mechanism for a pandas dataframe and plotting the resulting map with the original and privatized (lon, lat) coordinates and the radius.}
    \label{tab:functions_geoindis}
\end{center}

Coming back to Table~\ref{tab:functions_geoindis}, in the \texttt{metric\_privacy()} function the inputs are as follows: \textit{df} (the pandas dataframe with the data), \textit{column\_lat} (column with the latitude), \textit{column\_lon} (column with the longitude), epsilon (the privacy budget), \textit{new\_cols} (boolean value indicating whether two new columns will be added or whether the original columns will be replaced), \textit{earth\_radius\_m} (radius of the Earth, set to 6,371,000 meters) and \textit{seed} (random seed, by default 42). The output of this function is a new dataframe with new columns for latitude and longitude sanitized (or replacing the original ones) and a column with the calculated radius in each case (allowing users to inspect the magnitude of the perturbation associated with a chosen $\epsilon$).

To visually inspect the effect of the mechanism, \texttt{trasgoDP} also provides \texttt{plot\_metric\_dp\_map()}, which renders an interactive map comparing original and privatized locations, as illustrated below for the \textit{trip dataset}. This function allows to save the map in HTML format and has the following input values: \textit{df\_dp} (dataframe with the latitude and longitude columns privatized using the function \texttt{metric\_privacy()} with the attribute \textit{new\_cols} set to true), \textit{column\_lat} (original column with the latitude), \textit{column\_lon} (original column with the longitude) and \textit{save\_file} (path for saving the map in HTML format). 

One code example for using these two functions is given in Example Code~\ref{code:geo_indis}:

\begin{code}
\begin{minted}[fontsize=\small]{python}
import pandas as pd
from trasgodp.geoindis import metric_privacy, plot_metric_dp_map

# Read the data
data = pd.read_csv("trip_data.csv")
column_lat = "pickup_latitude"
column_lon = "pickup_longitude"

# Apply metric privacy creating new columns for lat and lon:
epsilon = 1.e-3
data_priv = metric_privacy(data, column_lat, column_lon, epsilon, new_cols=True)

# Plot and save the map:
plot_metric_dp_map(data_priv, column_lat, column_lon, save_file="example_map.html")

\end{minted}
\caption{Example: applying the geo-indistinguishability mechanism to the \textit{NYC taxis dataset} and plotting the resulting map.}
\label{code:geo_indis}
\end{code}
\vspace{0.3cm}

\subsection*{Privacy-utility trade-off metrics}

With the aim of providing users with metrics to quantify the quality of the noised data generated using DP, particularly in terms of distributional consistency, we have implemented specific functions in \texttt{trasgoDP}.

First, the most intuitive approach is to compute the divergence between the original column and the one obtained after applying DP, in order to quantify the information loss. To this end, different divergence metrics are calculated, including Total Variation Distance (TVD), Jensen-Shannon divergence (JS), and Kullback-Leibler divergence (KL).

In addition, we have defined a novel metric to quantify correlation loss (expressed as a percentage), which is based on measuring how well correlations between features are preserved. The idea is to assess how the correlation between a given column and a set of other features changes after applying differential privacy. To do this, a set of features is selected with respect to which the correlation is computed using a chosen method. Then, the absolute error between the original and perturbed correlations is calculated and transformed into a relative error expressed as a percentage.

In this way, large changes in correlations result in higher loss values. Since the metric is relative, it evaluates the error with respect to the ``strength'' of the original correlations. Thus, the procedure to compute the proposed metric is as follows:

\paragraph{Correlation loss (\%)}

\begin{itemize}
    \item[1.] Be $D \in \mathbb{R}^{n \times d}$ the original dataset and $D' \in \mathbb{R}^{n \times d+1}$ the privatized one (if one new column transformed with DP), or $D' \in \mathbb{R}^{n \times d}$ (if the privatized column has been substituted). Let's assume that $D' \in \mathbb{R}^{n \times d}$. Be $F = \{X_1, \dots, X_f\}$ the set of features selected, $f \leq d$.
    
    \item[2.] For each categorical feature $X_j \in F$, we define a function $\phi_j : \mathcal{C}_j \to \mathbb{Z}$ with $\mathcal{C}_j$ the set of values observed in $D$ and $D'$. Then we get the transformed datasets $\tilde{D}$ and $\tilde{D}'$.
    
    \item[3.] We extract the matrix based on the selected features: $X = \tilde{D}[F],$ $X' = \tilde{D}'[F]$.
    
    \item[4.] Be $\rho(\cdot,\cdot)$ a correlation method (Pearson, Spearman o Kendall). Then we calculate the correlation matrix as follows: $R = (\rho_{ij})_{i,j=1}^d$ with $\rho_{ij} = \rho(X_i, X_j)$ and $R' = (\rho'_{ij})_{i,j=1}^d$ with $\rho'_{ij} = \rho(X'_i, X'_j)$

    \item[5.] Remove autocorrelation: $\mathcal{I} = \{(i,j)\,:\, i \neq j,\; 1 \leq i,j \leq d\}$

    \item[6.] Difference between correlations: $\Delta_{ij} = |\rho_{ij} - \rho'_{ij}|, \quad (i,j) \in \mathcal{I}$.

    \item[7.] Mean of the difference between correlations: $\mu_{\Delta} = \frac{1}{|\mathcal{I}|} \sum_{(i,j)\in \mathcal{I}} |\rho_{ij} - \rho'_{ij}|$, and mean of the original correlation matrix
    $\mu_{R} = \frac{1}{|\mathcal{I}|} \sum_{(i,j)\in \mathcal{I}} |\rho_{ij}|$.

    \item[8.] Correlation loss (\%): $\mathcal{L} = 100 \cdot \frac{\mu_{\Delta}}{\mu_{R}}$.
\end{itemize}

The function for computing the correlation loss is available in the metrics subpackage, along with the functions for computing divergences. In the latter case, it is sufficient to provide the original and differentially private datasets, the sanitized column, and whether it has been replaced or newly created.
In contrast, for the correlation loss, instead of specifying a single sanitized column, we provide the set of features over which we want to measure the correlation. In both cases, it is necessary to indicate whether the sanitized data has replaced the original column or has been added as a new one. A battery of examples on how to use this novel function, along with guidance on their interpretation, are presented below.

\subsubsection*{Tests and examples}

\paragraph{\textit{Adult dataset}}
For conducting the test of the library, the \textit{adult dataset} was used. In addition, we also use such dataset for illustrating the correlation loss function. First, in Table~\ref{tab:adult_extraction} we show an extraction of ten rows and eight columns that have been used during the testing phase.

\begin{center}
    \resizebox{\linewidth}{!}{
    \begin{tabular}{cccccccc}
    \toprule
    \textit{\textbf{workclass}} & \textit{\textbf{education}} & \textit{\textbf{marital-status}} & \textit{\textbf{occupation}} & \textit{\textbf{sex}} & \textit{\textbf{native-country}} & \textit{\textbf{age}} & \textit{\textbf{salary-class}}\\
    \midrule
    Private & 11th  & Never-married & Machine-op-inspct & Male & Puerto-Rico & 48 & <=50K\\
    Private & Bachelors & Never-married & Exec-managerial & Male & Germany & 30 & <=50K\\
    Self-emp-not-inc & HS-grad & Married-civ-spouse & Other-service & Male & Canada & 56 & >50K\\
    Local-gov & Some-college & Divorced & Adm-clerical & Female & Mexico & 47 & <=50K\\
    Self-emp-not-inc & HS-grad & Married-civ-spouse & Farming-fishing & Male & Cambodia & 42 & >50K\\
    Self-emp-not-inc & 9th & Married-civ-spouse & Craft-repair & Male & Portugal & 26 & <=50K\\
    Self-emp-inc & 5th-6th & Married-civ-spouse & Transport-moving & Male & Cuba & 47 & <=50K\\
    Local-gov & Bachelors & Divorced & Adm-clerical & Female & Honduras & 36 & <=50K\\
    Self-emp-not-inc & HS-grad & Married-civ-spouse & Other-service & Female & Italy & 47 & <=50K\\
    Private & Bachelors & Never-married & Exec-managerial & Female & United-States & 49 & >50K\\
    \bottomrule
    \end{tabular}}
    \captionof{table}{Sample rows and columns extracted from the \textit{adult dataset} with a selected subset of columns that can act as quasi-identifiers and sensitive attribute for data privatization tasks.}
    \label{tab:adult_extraction}
\end{center}

\begin{figure}[pos=htbp]
	\centering
    \subfigure[Laplace mechanism. Column: Age.]{
		\includegraphics[width=0.45\textwidth]{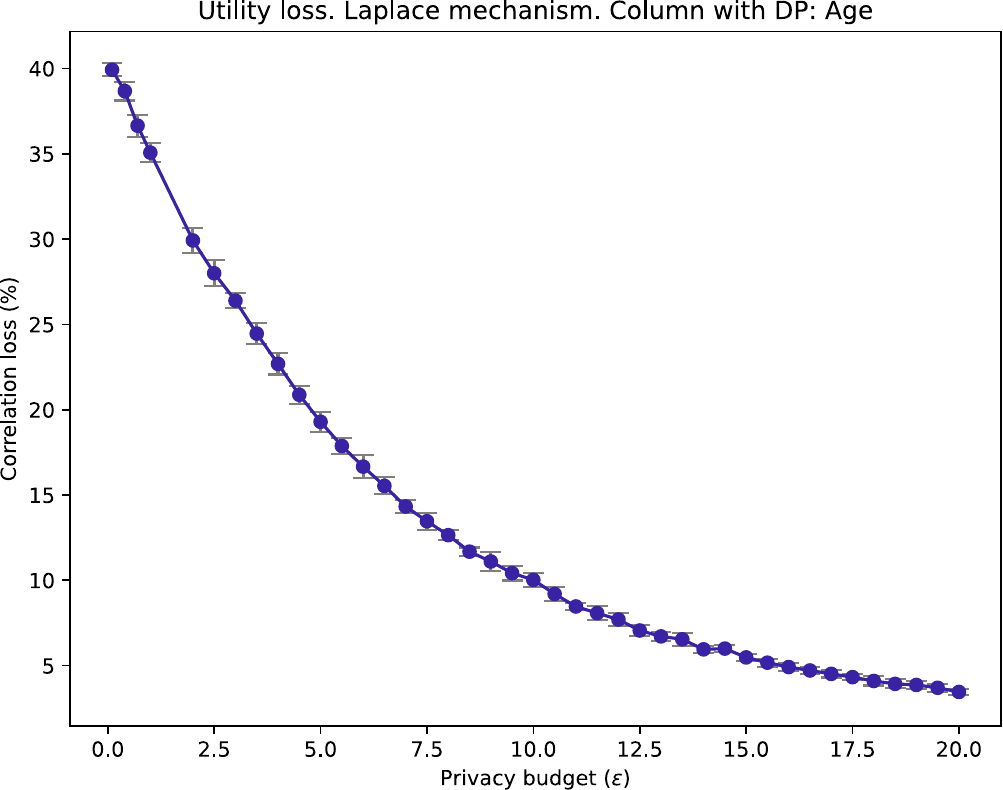}
		\label{fig:laplace_age_adult}
	}
    \hfill 
    \subfigure[Gaussian mechanism. Column: Age.]{
		\includegraphics[width=0.45\textwidth]{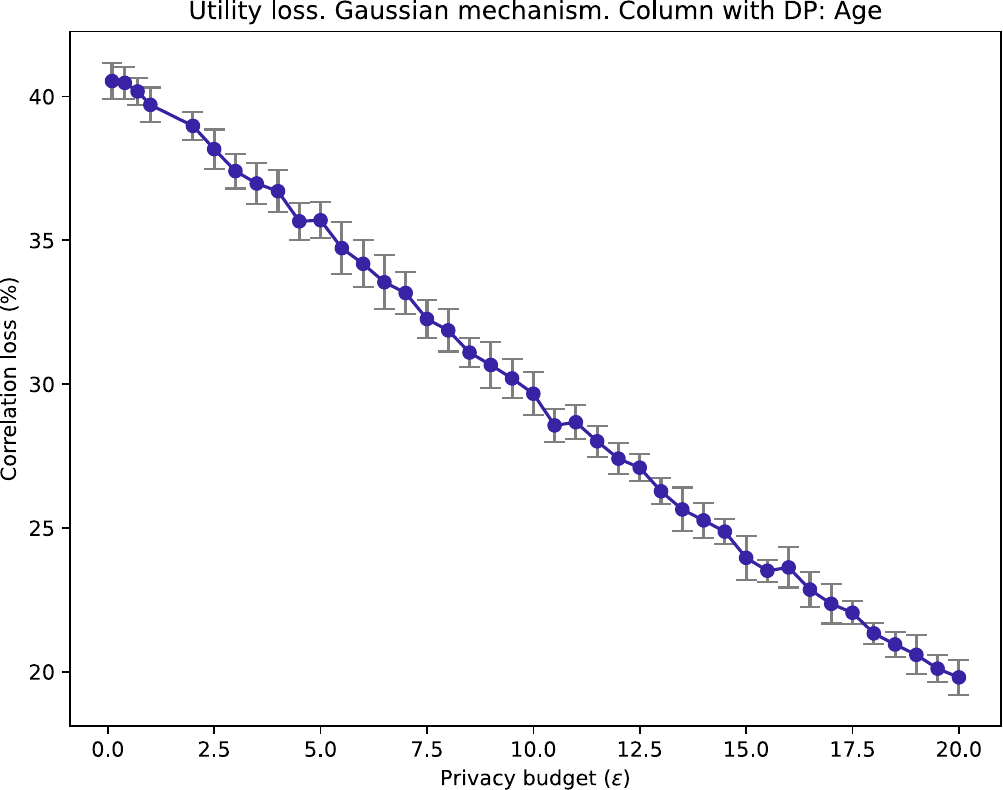}
		\label{fig:gauss_age_adult}
	}
    \vfill
    \subfigure[Exponential mechanism. Column: Workclass.]{
		\includegraphics[width=0.45\textwidth]{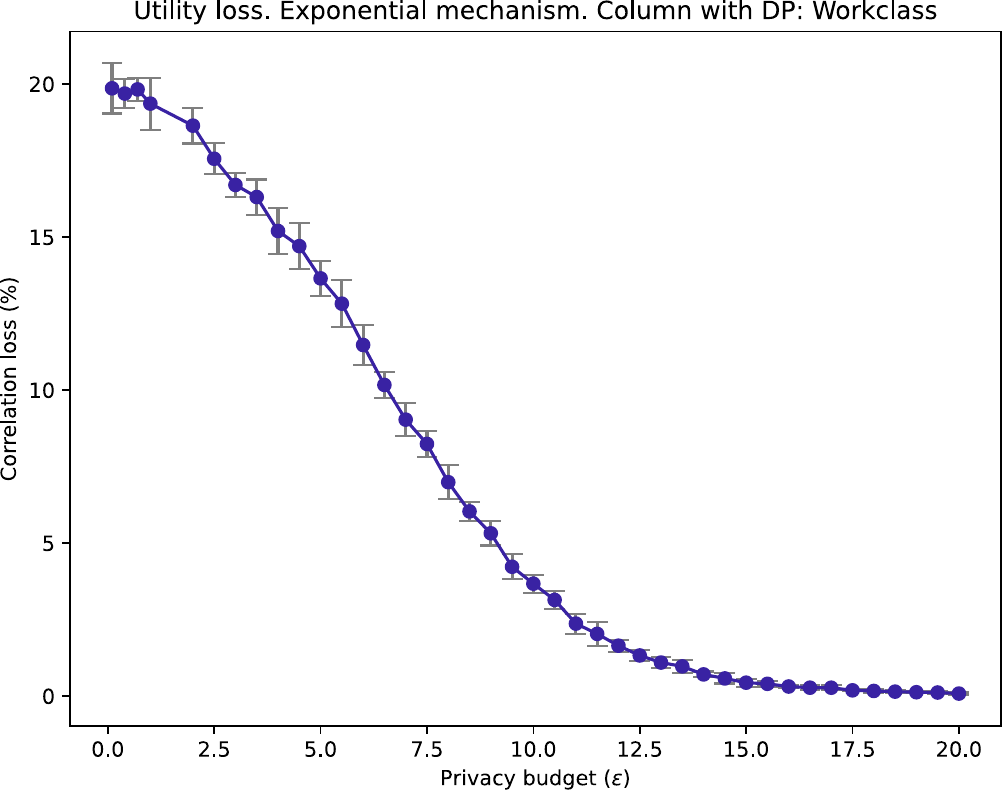}
		\label{fig:exp_workclass}
	}
    \hfill
    \subfigure[k-ary randomized response mechanism. Column: Workclass.]{
		\includegraphics[width=0.45\textwidth]{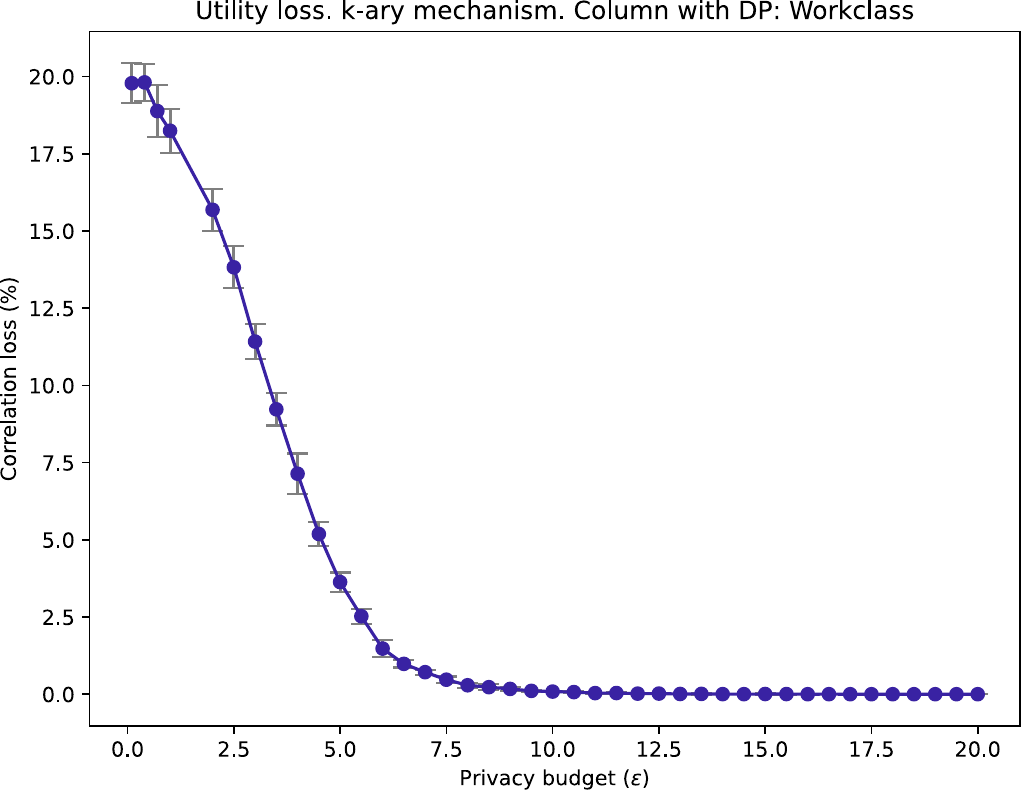}
		\label{fig:kary_workclass}
	}
    \vfill
    \subfigure[Exponential mechanism. Column: Sex.]{
		\includegraphics[width=0.45\textwidth]{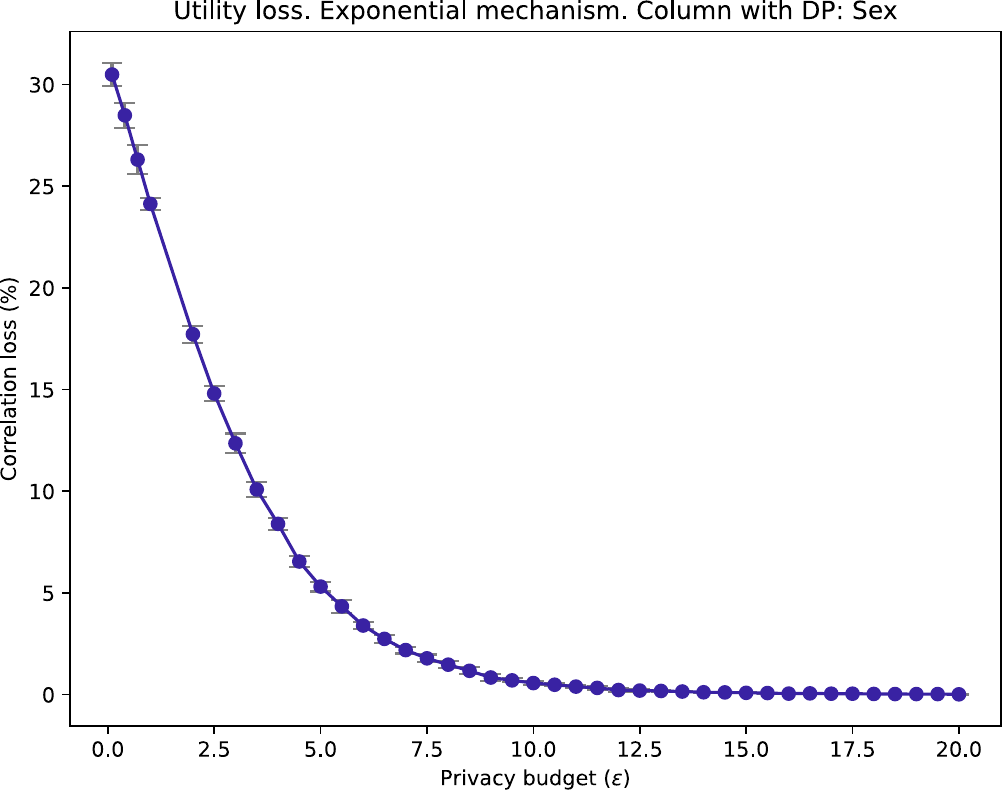}
		\label{fig:exp_bin_sex}
	}
    \hfill
    \subfigure[Binary randomized response mechanism. Column: Sex.]{
		\includegraphics[width=0.45\textwidth]{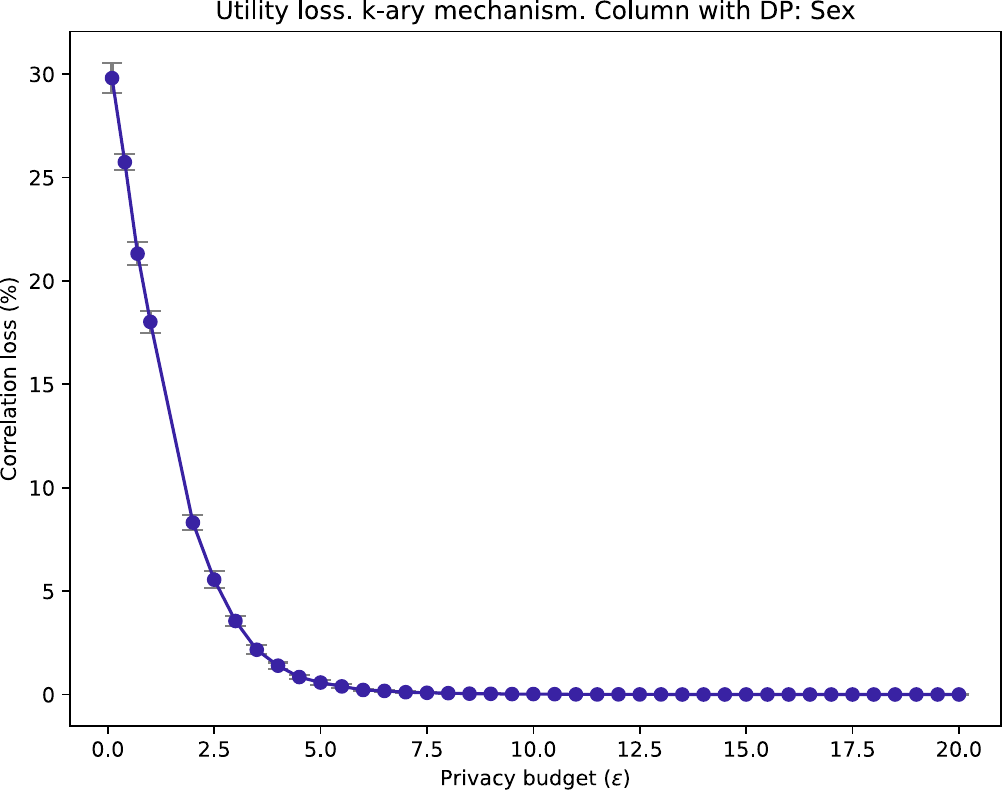}
		\label{fig:kary_bin_sex}
	}
	\caption{Utility loss calculated when varying $\epsilon$ during 10 runs, based on the differences between the correlation matrix. Comparison between mechanisms for numerical and categorical (binary and non-binary) attributes. \textit{Adult dataset}.}
	\label{fig:corr_adult}
\end{figure}

In order to calculate the correlation loss and measure the impact of different values of the privacy budget, the results obtained for a numerical column (\textit{age}) are shown below, using both the Laplace and Gaussian mechanisms; for a non-binary categorical column (\textit{workclass}), using the Exponential and Randomized Response mechanisms; and finally for a binary categorical column (\textit{sex}), using both the Exponential and Binary Randomized Response mechanisms. Specifically, the following set of $\epsilon$ values were evaluated: $\epsilon \in \{0.1 + k\cdot 0.3 | k\in\{0, 1, 2, 3\}\}\cup \{2 + k\cdot 0.5 | k\in\mathbb{Z}, 1 \leq k \leq 20\}$, with $\delta= 10^{-6}$ in the case of the Gaussian mechanism. For the \textit{age}, 16 and 100 were used as lower and upper bounds respectively. Concerning the features used for computing the correlation, all the columns presented in Table~\ref{tab:adult_extraction} were used. Finally, for each value of $\epsilon$, the procedure was repeated 10 runs to obtain the mean and standard deviation for each case. These results are shown graphically in Figure~\ref{fig:corr_adult} for each variable and mechanism applied.

From the graphs in Figure~\ref{fig:corr_adult}, it is immediately apparent that, as expected, a smaller value of $\epsilon$ increases the correlation loss, while a larger value substantially reduces it. Regarding the comparison of numerical mechanisms, it is observed that the flexibility provided by the delta parameter (by allowing the privacy budget to be exceeded with a certain probability) results in a correlation loss up to three times greater than with the Laplace mechanism for the same value of $\epsilon$. Thus, the curve corresponding to the Gaussian mechanism follows an approximately linear decreasing trend, while that of Laplace decreases exponentially, converging more rapidly toward low loss values.

As for the results for the non-binary categorical variable (\textit{workclass}), the Exponential mechanism exhibits a slower reduction in correlation loss than the k-ary Randomized Response, with both following an inverse sigmoid curve, although with different slopes. This behavior is also observed for the binary categorical variable (\textit{sex}), where the correlation is close to zero with a privacy budget of approximately $\epsilon = 5$ for the k-ary mechanism and $\epsilon = 10$ for the Exponential mechanism. It is immediately noticeable how categorical variables, by taking a much smaller range of values, require a lower value of $\epsilon$ to maintain a reasonable trade-off between privacy and utility in terms of correlation, while numerical variables covering a wider range of values, such as age, require higher values of $\epsilon$ for the perturbed distribution to resemble the original.

\paragraph{\textit{Global cancer patients dataset}}
We will now consider a second example, based on the application of DP mechanisms implemented to the global cancer patients dataset\cite{global_cancer_patients_2015_2024}, which contains synthetic global cancer patient data reported from 2015 to 2024, including demographic variables such as age, gender, and country, as well as genetic risk, cancer type, and stage, in addition to treatment cost, survival rate, and severity. Even if this dataset doesn't contain real-world patient data, it is inspired by statistics from the World Health Organization (WHO), data from the National Institutes of Health (NIH) and trends extracted from the Global Cancer Observatory (GCO) and Our World in Data. Table~\ref{tab:disease_extraction} shows an extract of 10 rows and the 10 columns mentioned above, selected from among the 15 columns and 50000 rows available. Note that this dataset is used here only as an illustrative example to demonstrate the applicability of the implemented mechanisms and metrics across two datasets with different statistical properties.

\begin{center}
    \resizebox{\linewidth}{!}{
    \begin{tabular}{cccccccccc}
    \toprule
    \textit{\textbf{Age}} & \textit{\textbf{Gender}} & \textit{\textbf{Country\_Region}} & \textit{\textbf{Year}} & \textit{\textbf{Genetic\_Risk}} & \textit{\textbf{Cancer\_Type}} & 	\textit{\textbf{Cancer\_Stage}} & \textit{\textbf{Treatment\_Cost\_USD}} & \textit{\textbf{Survival\_Years}} & \textit{\textbf{Target\_Severity\_Score}} \\
    \midrule
    71 & Male & UK & 2021 & 6.4 & Lung & Stage III & 62913.44 & 5.9 & 4.92\\
    34 & Male & China & 2021 & 1.3 & Leukemia & Stage 0 & 12573.41 & 4.7 & 4.65\\
    43 & Female & Brazil & 2017 & 5.1 & Skin & Stage III & 77977.12 & 2.9 & 3.62\\
    40 & Female & USA & 2018 & 6.4 & Colon & Stage I & 49619.66 & 0.4 & 6.03\\
    41 & Male & Canada & 2021 & 5.1 & Cervical & Stage 0 & 9790.83 & 1.0 & 5.05\\
    85 & Other & India & 2016 & 2.8 & Leukemia & Stage III & 17158.84 & 3.0 & 4.55\\
    83 & Other & Russia & 2023 & 3.2 & Prostate & Stage II & 38290.91 & 1.1 & 4.09\\
    86 & Male & Australia & 2015 & 8.6 & Skin & Stage 0 & 98333.43 & 0.1 & 4.15\\
    22 & Male & Germany & 2018 & 9.5 & Cervical & Stage IV & 33468.99 & 9.5 & 5.98\\
    67 & Female & China & 2015 & 2.6 & Liver & Stage II & 65060.21 & 9.7 & 3.85\\
    \bottomrule
    \end{tabular}}
    \captionof{table}{Sample rows and columns extracted from the \textit{global cancer dataset} with a selected subset of columns that can act as quasi-identifiers and sensitive attribute for data privatization tasks.}
    \label{tab:disease_extraction}
\end{center}

\begin{figure}[pos=htbp]
	\centering
    \subfigure[Laplace mechanism. Column: Age.]{
		\includegraphics[width=0.43\textwidth]{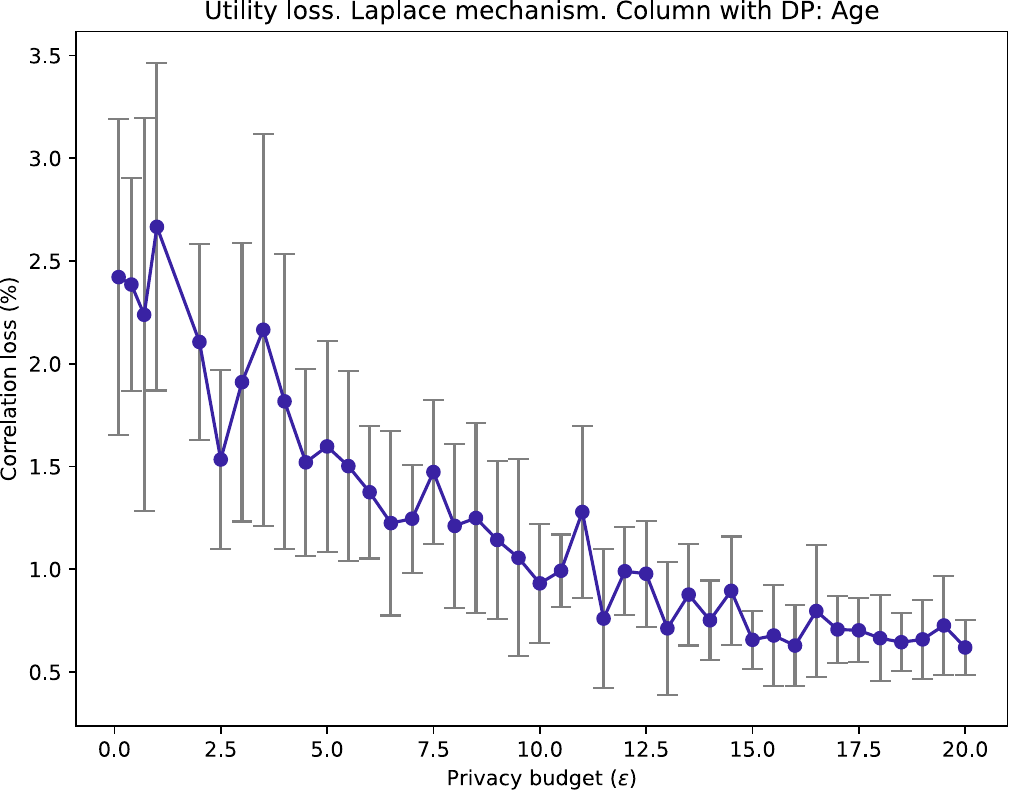}
		\label{fig:laplace_age_patients}
	}
    \hfill
    \subfigure[Gaussian mechanism. Column: Age.]{
		\includegraphics[width=0.43\textwidth]{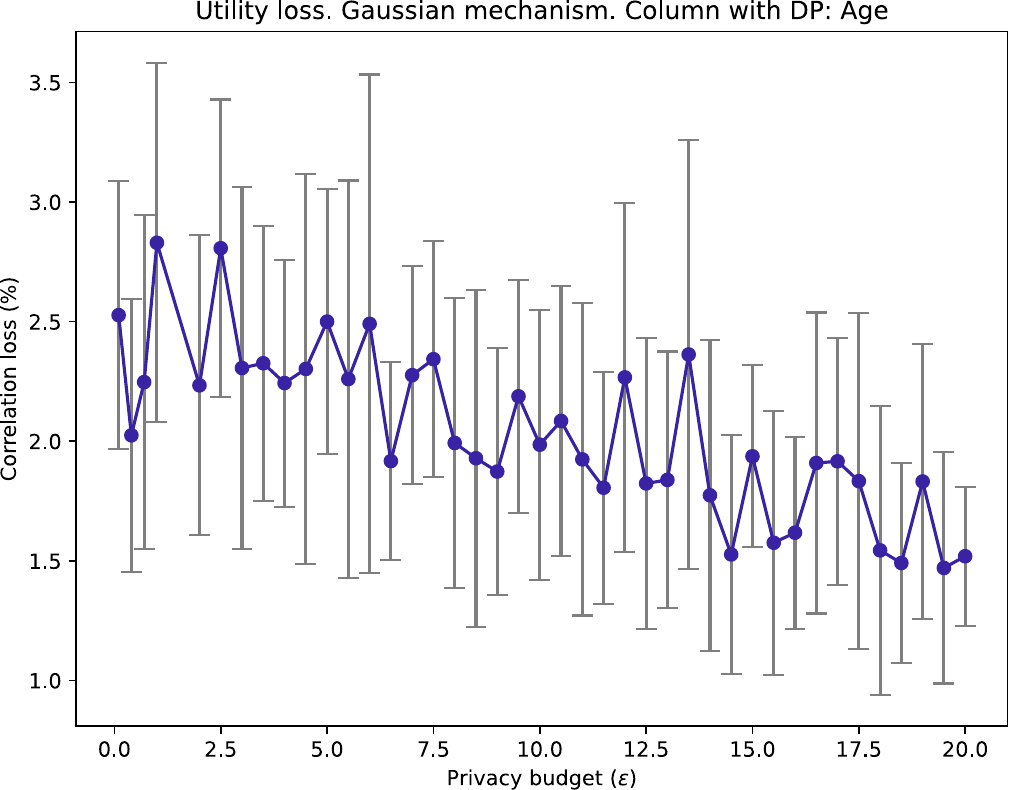}
		\label{fig:gauss_age_patients}
	}
    \vfill
    \subfigure[Exponential mechanism. Column: Country\_Region.]{
		\includegraphics[width=0.43\textwidth]{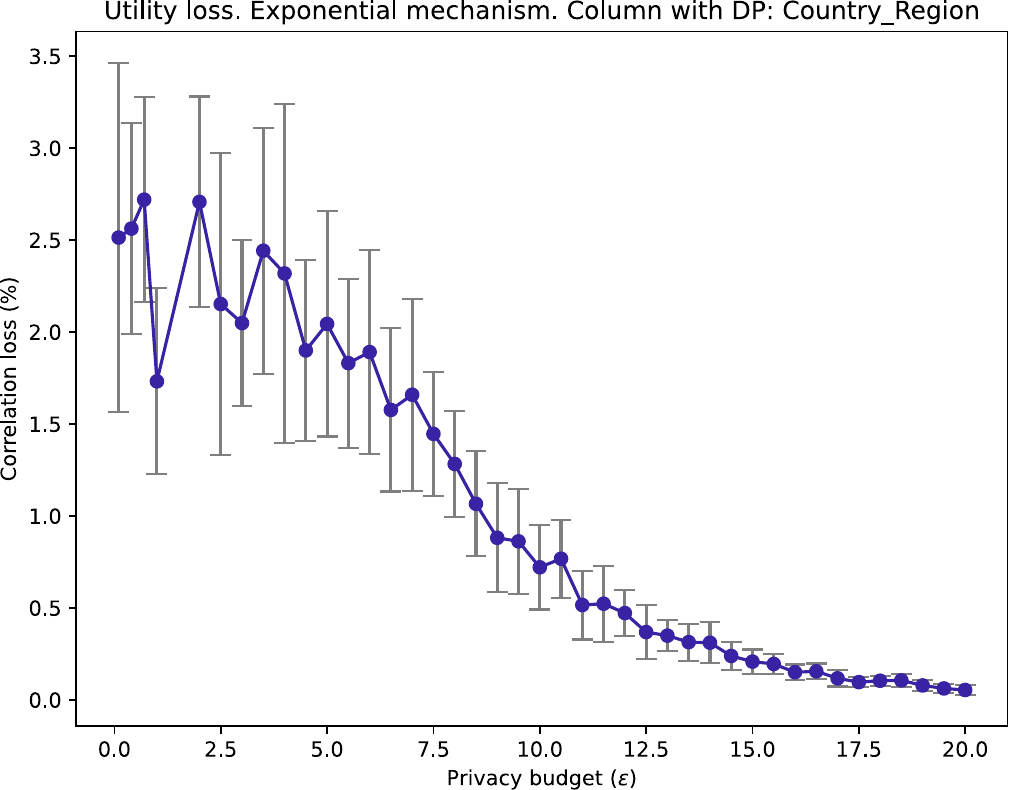}
		\label{fig:exp_country_patients}
	}
    \hfill
    \subfigure[k-ary randomized response mechanism. Column: Country\_Region.]{
		\includegraphics[width=0.43\textwidth]{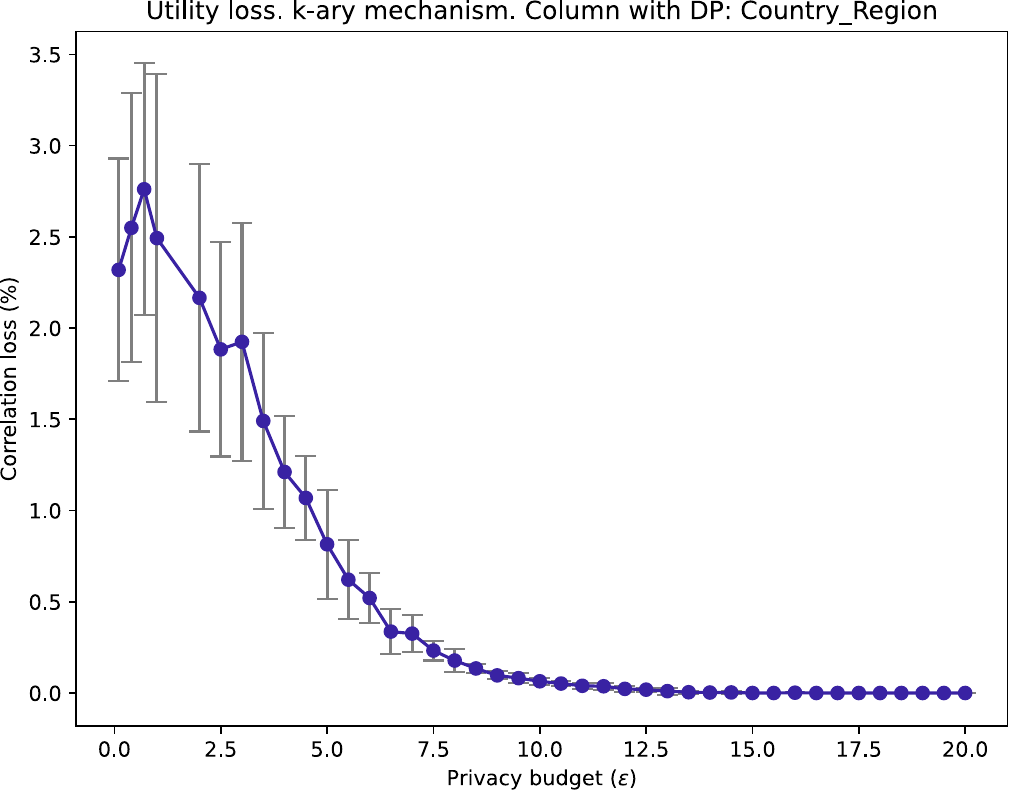}
		\label{fig:kary_country_patients}
	}
	\caption{Utility loss calculated when varying $\epsilon$ during 10 runs, based on the differences between the correlation matrix. Comparison between mechanisms for numerical and categorical attributes. Global patients dataset.}
	\label{fig:corr_global_patients}
\end{figure}

In this case, there are no binary variables available (the gender variable takes on three possible values), but we present the results for the numerical variable age, as well as the categorical variable country. Note that in both this example and the previous one, we focus on this kind of demographic variables rather than on sensitive attributes, since a priori, what we want is to protect against the possibility of identifying an individual in the database; therefore, we focus specifically on the QIs. Thus, Figure~\ref{fig:corr_global_patients} shows the results obtained for these two variables using the Laplace and Gaussian mechanisms for age (again using 16 and 100 as lower and upper bounds) and the exponential and k-ary mechanisms for country. Concerning the features used for computing the correlation loss, in addition to the feature transformed with DP in each case, the following were used in this example: \textit{Genetic\_Risk}, \textit{Cancer\_Type}, \textit{Cancer\_Stage}, \textit{Treatment\_Cost\_USD}, \textit{Survival\_Years} and \textit{Target\_Severity\_Score}. Note that, once again, the same privacy budget values are used as in the previous example, and the experiments are repeated 10 runs in each case.

In this case, the loss of correlation is significantly smaller than in the previous cases, although the mechanisms follow a similar trend curve: Laplace decreases more quickly than Gaussian, and k-ary more than exponential.

\begin{figure}[pos=htbp]
    \centering
    \includegraphics[width=0.8\linewidth]{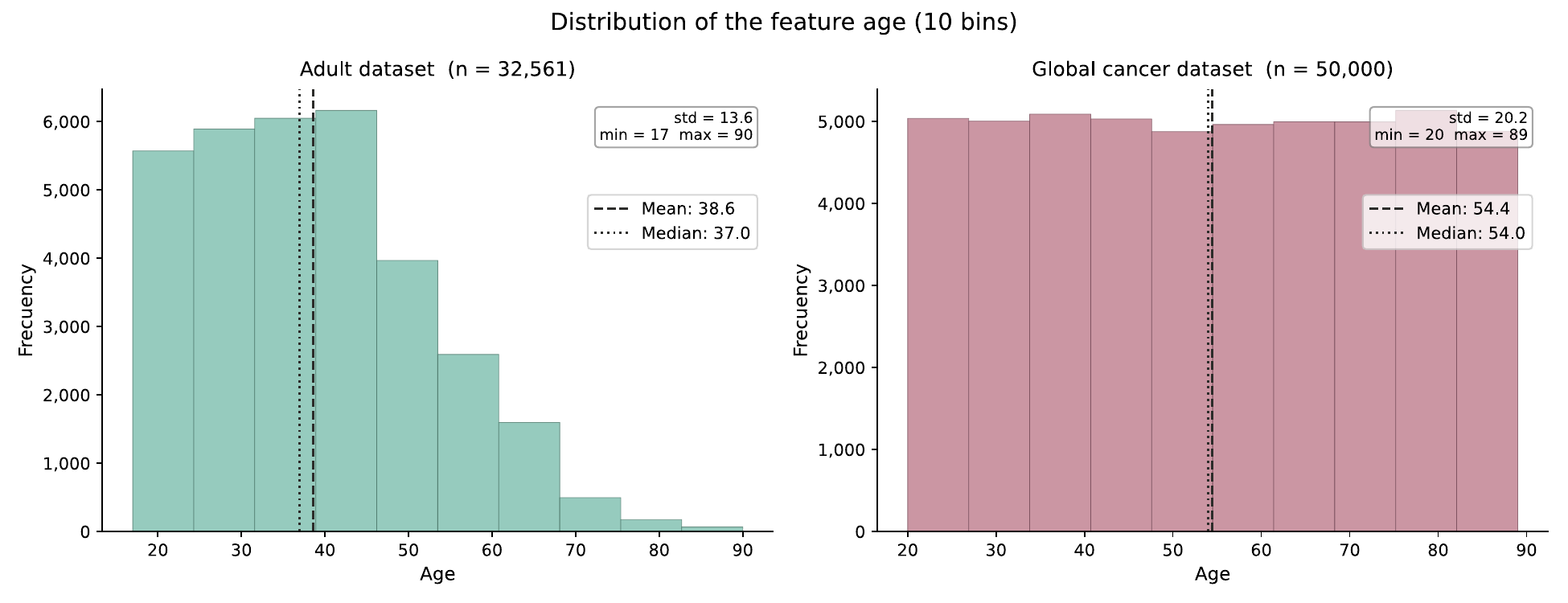}
    \caption{Distribution of the age feature in the \textit{adult dataset} and in the \textit{global cancer dataset} (10 bins).}
    \label{fig:age_10bins}
\end{figure}

Regarding the differences in the results obtained for the same epsilon values with both datasets, it is important to start by analyzing the distributions of the variables in each case. As for the number of unique values, the numbers are very similar: while the \textit{adult dataset} had 73 unique values for \textit{age}, the cancer dataset contains 70 unique values. For the \textit{country} variable, this dataset has 10 unique values, compared to the 9 unique values for \textit{workclass} in adult. With regard to column \textit{sex} in the adult dataset, the value \textit{'Male'} appears in the 66.92$\%$ of the cases times, compared to \textit{'Female'} in the 33.08$\%$.

However, the most significant differences are found in the distributions of the variables. Starting with age, Figure~\ref{fig:age_10bins} shows the age distribution in both cases-approximately uniform in the \textit{global cancer dataset} and unimodal with moderate positive skewness (right-skewed) in the \textit{adult dataset}.

\begin{figure}[pos=htbp]
    \centering
    \includegraphics[width=0.8\linewidth]{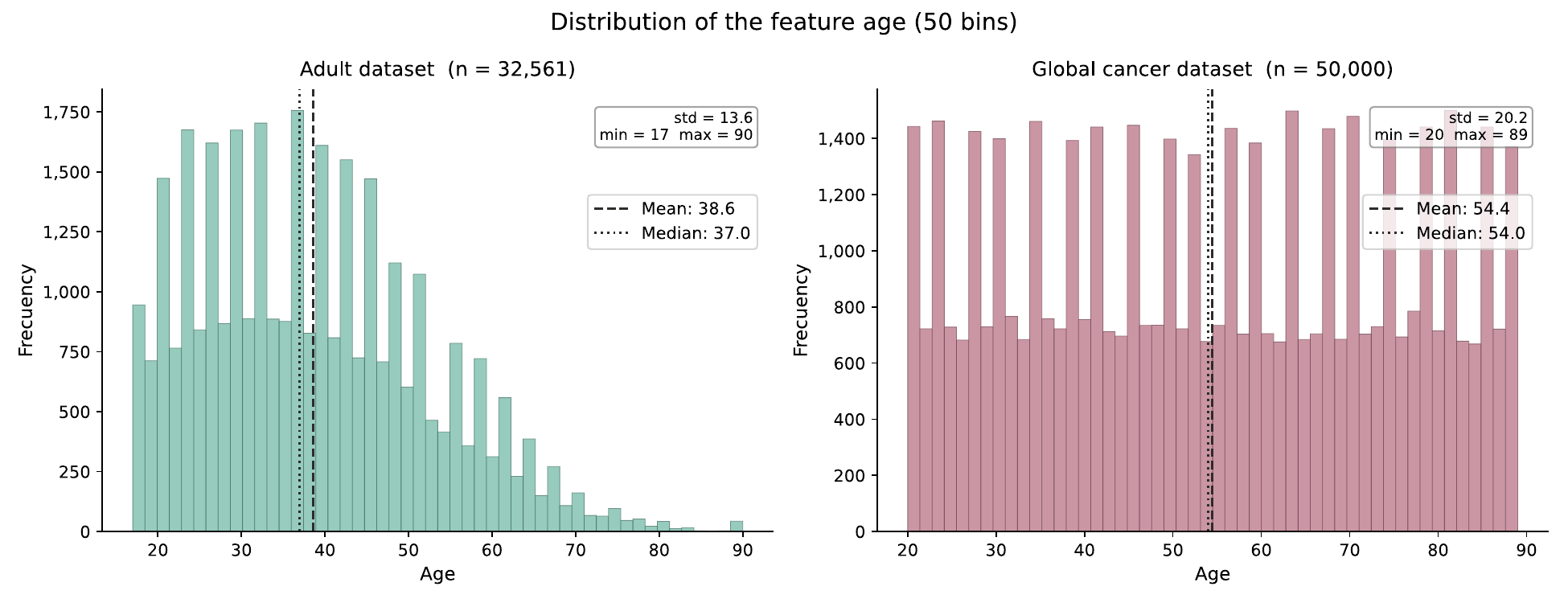}
    \caption{Distribution of the age feature in the \textit{adult dataset} and in the \textit{global cancer dataset} (50 bins).}
    \label{fig:age_50bins}
\end{figure}

\begin{figure}[pos=htbp]
    \centering
    \includegraphics[width=0.8\linewidth]{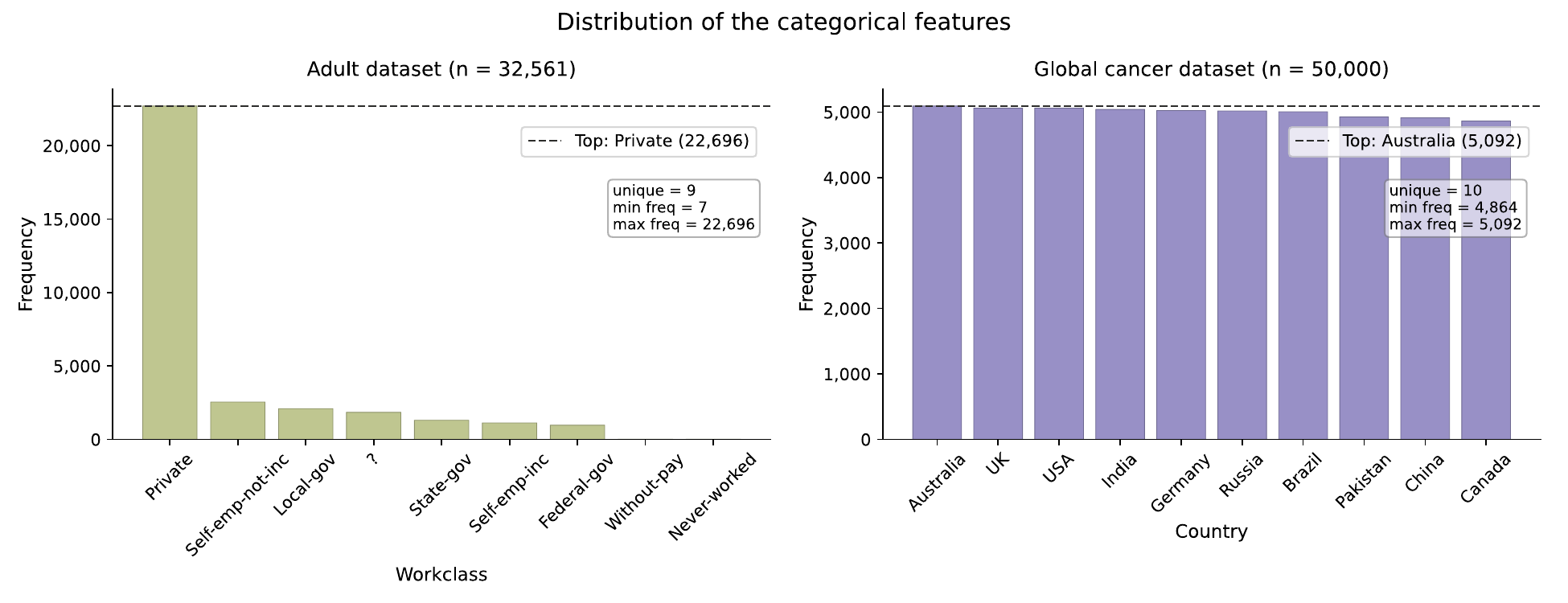}
    \caption{Distribution of the non-binary categorical features analyzed in the \textit{adult dataset} (\textit{workclass}) and in the \textit{global cancer dataset} (\textit{country}).}
    \label{fig:barplot_cat}
\end{figure}

Specifically, applying a $\chi^2$ test to the age variable in the \textit{global cancer dataset} comparing the observed frequencies with the expected frequencies under a uniform distribution (5,000 values per bin in the case of 10 bins), yields a p-value of 0.217, which does not allow us to reject the null hypothesis that the variable follows a uniform distribution. In contrast, in the case of adult, a normal test using the D'Agostino and Pearson's tests confirms that the variable does not follow a normal distribution (with a p-value=0, as expected); given a skewness of 0.559, it is concluded that the distribution presents moderate positive skewness. The previous is verified when using 10 bins, as shown in Figure~\ref{fig:age_10bins}. However, Figure~\ref{fig:age_50bins} shows the results obtained when considering 50 bins in each case. Under that approach, we can note (and verify with a hypothesis test), that for the \textit{global cancer dataset}, the age doesn't follow a uniform distribution.

In addition, Figure~\ref{fig:barplot_cat}, presents bar charts showing the distributions of the \textit{workclass} and \textit{country} variables. While the variable studied in the \textit{adult dataset} has a very different distribution for each possible value, the variable studied in the \textit{global cancer dataset} exhibits a distribution that follows a uniform one. The impact of these distributions is clearly reflected in Figures~\ref{fig:corr_adult} and \ref{fig:corr_global_patients}, with the results obtained for the correlation loss in the the \textit{global cancer dataset} being smaller than the ones for the \textit{adult dataset} under a similar number of unique values and same noise multipliers.

Concerning the use of this metric and calculation of the divergence between the distribution of a column in the original and DP datasets, two functions are available in the \texttt{trasgodp.metrics} package: \texttt{correlation\_loss()} and \texttt{divergence\_distributions()}. The first one returns a float with the utility loss ($\%$), and the second one returns a dictionary with the divergence metrics (TVD, JS, KL). Concerning the inputs, both require the original dataframe, the one with DP, the column to be analyzed and if a new column was created when applying DP. In addition, for the correlation loss the list of features to be included in the correlation analysis and the method used (Pearson, Kendall, or Spearman). Example Code~\ref{code:corr_loss_global} shows how to perform this calculation for the column \textit{age} of the \textit{global cancer dataset}, given the set of features used in the previous example.

\begin{code}
\begin{minted}[fontsize=\small]{python}
import pandas as pd
import numpy as np
from trasgodp import numerical, metrics

# Read and process the data
df_original = pd.read_csv("global_cancer_patients_2015_2024.csv")

# Select the features to be used when computing the correlation
features = [
    "Genetic_Risk",
    "Cancer_Type",
    "Cancer_Stage",
    "Treatment_Cost_USD",
    "Survival_Years",
    "Target_Severity_Score"
]

# Apply DP with the Laplace mechanism
column = 'Age'
epsilon = 1
df_dp = numerical.dp_clip_laplace(
    df_original, column, epsilon, lower_bound = 16, upper_bound = 100, new_column=True
)

# Get the correlation loss for the column 'Age':
features_col = features + [column]
corr_loss = metrics.correlation_loss(
    df_original, 
    df_dp, 
    features_col, 
    new_column=True
)

print(f'Correlation loss (%) for {column}: {np.round(corr_loss, 3)}')
\end{minted}
\caption{Example: calculating the correlation loss in a particular run for the column \textit{age} of the \textit{global cancer dataset}.}
\label{code:corr_loss_global}
\end{code}

\vspace{0.1cm}

\paragraph{\textit{NYC taxis trip dataset}}

Finally, to test the geo-indistinguishability mechanism and assess the metric-privacy guarantees provided by \texttt{trasgoDP}, we use a publicly available dataset containing information about the New York City taxi trips through 2013 \cite{nyc_taxis}. This dataset gained popularity in the privacy research community after it was demonstrated that the pseudonymization applied to the original version (hashing taxi license and medallion numbers using the MD5 function) could be easily reversed, since the space of valid license numbers is small enough to be exhaustively searched, allowing for the 
complete re-identification of each individual trip. 

In addition, based on the de-anonymized dataset, different works cross-referenced the 
pickup and drop-off locations with time-stamped photographs of celebrities getting into or out of taxis in New York City, successfully inferring the exact fares for their trips, the tips, and, in several cases, their destinations, which are strongly correlated with their home or work addresses. This case illustrates how location microdata that seem harmless or that only allow only limited identification on their own, can reveal a great amount of sensitive information when linked to publicly available auxiliary information \cite{7796899}. This reinforces the need for formal privacy guarantees, such as those provided by geo-indistinguishability.

Specifically, a version of this data, available in \href{https://github.com/IFCA-Advanced-Computing/trasgoDP/tree/main/examples}{the examples folder of the \texttt{trasgoDP} GitHub repository} has been used. It contains 100,000 records, including information such as the latitude and longitude coordinates of departure and arrival, the duration, distance and date of the trip, the pick-up and drop-off time, number of passengers etc. Table~\ref{tab:taxis_extraction} shows an extraction of the first ten rows, restricted to the subset of columns mentioned above and only departure coordinates for the sake of simplicity.

\begin{figure}[pos=htbp]
    \centering
    \includegraphics[width=0.9\linewidth]{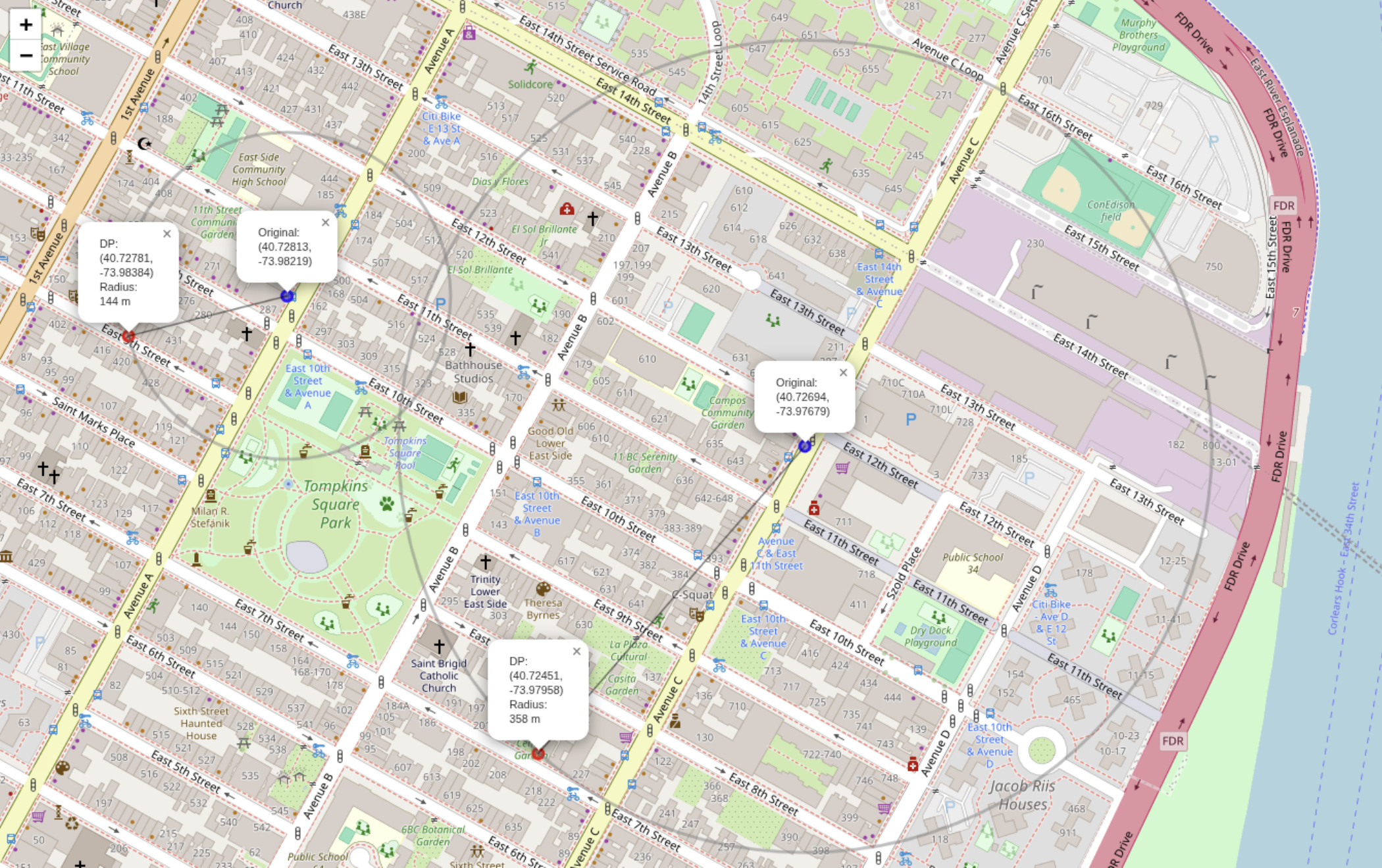}
    \caption{Example of a specific point (lat, lon) within the NYC taxis trip dataset, showing the original value and the value obtained by applying the geo-indistinguishability mechanism.}
    \label{fig:geo_indis_map_example}
\end{figure}

Given this dataset, the geo-indistinguishability mechanism is applied to privatize the latitude and longitude values of the pick up, producing a new pair of coordinates for each record together with the corresponding perturbation radius. In addition, as already explained as shown in Example Code~\ref{code:geo_indis}, users can render the resulting map as an interactive HTML file, which can be saved locally, allowing a direct visual comparison between the original and privatized locations. Figure~\ref{fig:geo_indis_map_example} shows a screenshot of the map with an example of how one of the original longitude and latitude pick-up points has been perturbed using the algorithm presented in the Example Code~\ref{code:corr_loss_global} for this dataset.

\begin{center}
    \resizebox{\linewidth}{!}{
    \begin{tabular}{ccccccc}
    \toprule
    \textit{\textbf{pickup\_datetime}} & \textit{\textbf{dropoff\_datetime}} & \textit{\textbf{passenger\_count}} & \textit{\textbf{trip\_time\_in\_secs}} & \textit{\textbf{trip\_distance}} & \textit{\textbf{pickup\_longitude}} & \textit{\textbf{pickup\_latitude}}\\
    \midrule    
    2013-12-05 22:45:00 & 2013-12-05 22:54:00 & 4 & 540 & 2.08 & -73.973854 & 40.762615 \\
    2013-12-05 22:53:00 & 2013-12-05 22:57:00 & 6 & 240 & 0.56 & -73.982300 & 40.766106 \\
    2013-12-05 22:51:00 & 2013-12-05 22:55:00 & 1 & 240 & 1.44 & -73.989204 & 40.757675 \\
    2013-12-05 22:39:00 & 2013-12-05 22:55:00 & 1 & 960 & 2.03 & -73.974663 & 40.763454 \\
    2013-12-05 22:54:00 & 2013-12-05 22:58:00 & 1 & 240 & 0.13 & -73.985802 & 40.722168 \\
    2013-12-03 13:18:00 & 2013-12-03 13:31:00 & 1 & 780 & 1.21 & -73.990067 & 40.745975 \\
    2013-12-03 13:16:00 & 2013-12-03 13:29:00 & 1 & 780 & 3.43 & -74.014221 & 40.709648 \\
    2013-12-03 13:25:00 & 2013-12-03 13:34:00 & 1 & 540 & 4.25 & -73.972458 & 40.746586 \\
    2013-12-05 23:49:00 & 2013-12-05 23:55:00 & 1 & 360 & 2.10 & -73.983429 & 40.773666 \\
    2013-12-03 14:22:00 & 2013-12-03 14:37:00 & 1 & 900 & 1.69 & -73.994888 & 40.750156 \\
    \bottomrule
    \end{tabular}}
    \captionof{table}{Sample rows and columns extracted from the \textit{NYC taxis trip dataset} with a selected subset of columns, including the latitude and longitude parameters of the pickup.}
    \label{tab:taxis_extraction}
\end{center}

\begin{figure}[pos=htbp]
    \centering
    \includegraphics[width=0.8\linewidth]{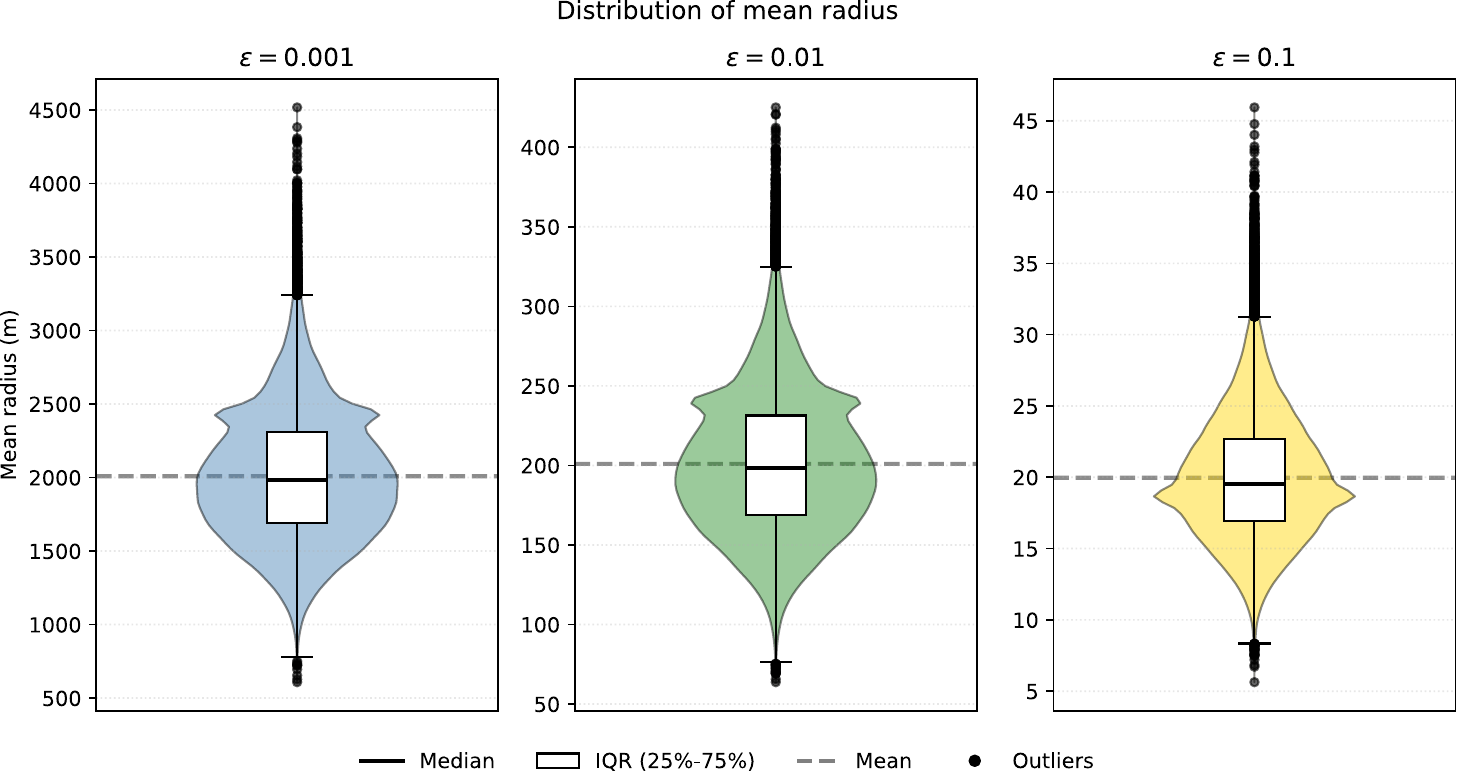}
    \caption{Distribution of the radius obtained for each record, mean after 10 runs with different seeds and with $\epsilon \in \{0.001, 0.01, 0.1\}$.}
    \label{fig:radius_taxi}
\end{figure}

Finally, for this third dataset, the geo-indistinguishability mechanism was run 10 times for each of three privacy budget values: $\epsilon \in \{0.001, 0.01, 0.1\}$. Figure~\ref{fig:radius_taxi} shows the distribution of the mean perturbation radius obtained for each record  across the 10 runs, for each of the three $\epsilon$ values considered. As expected, the perturbation radius decreases as $\epsilon$ increases, since larger values of $\epsilon$ correspond to weaker privacy guarantees and therefore less noise being added to the original coordinates.

This result illustrates the practical importance of carefully calibrating the privacy budget for this type of data: on the one hand, an excessively small value of $\epsilon$ can result in perturbation ranges of several kilometers, which would seriously compromise the usefulness of the published data for tasks such as identifying areas with high demand for pickups at different times of the day; on the other hand, an $\epsilon$ that does not provide sufficient protection may leave pickup locations close enough to their original coordinates that, as discussed above, they can still be used to infer sensitive information (such as the home or work address of an individual) especially when cross-referenced with external data sources. Therefore, selecting an appropriate value for $\epsilon$ requires striking a balance between these two opposing objectives.

\section*{DISCUSSION}

\subsection*{Conclusion: differential privacy in data science workflows}

When designing a data science workflow from zero, it is often assumed that the data lifecycle is in a controlled environment. However, when privacy constraints exist, it is necessary to incorporate additional mechanisms to prevent the extraction of sensitive information, whether directly from the data or indirectly through the trained models. When focusing on the privacy of the data, the first step typically involves removing or pseudonymizing identifiers, depending on whether they may need to be reconstructed in the future. Next, two main approaches can be considered: (1) first, anonymization through generalization, where QIs are transformed using generalization based on hierarchies to satisfy formal privacy definitions, such as $k$-anonymity, $\ell$-diversity, $t$-closeness, $\delta$-disclosure, etc. In this context, tools like \texttt{anjana} enable the anonymization process, while \texttt{pyCANON} facilitates the evaluation of the level of privacy achieved under different definitions. Both software products have been developed in an analogous way to \texttt{trasgoDP}, allowing for their easy integration within the same workflow. (2) Second, we may consider generating noisy data (following the idea of synthetic data generation but exploiting the initial raw data) using mechanisms based on local differential privacy or metric privacy, enabling microdata publication and thus analysis, while maintaining formal privacy guarantees. This is precisely the approach that \texttt{trasgoDP} aims to facilitate to incorporate in a complete data science pipeline.

Both strategies have advantages and limitations. On the one hand, anonymization based on generalization can significantly reduce the usefulness of the data due to a loss of granularity; for example, an exact age provides more information than an age range. However, this approach accurately preserves the overall distribution of the original dataset within a generalization domain. On the other hand, LDP mechanisms maintain the original domain of the variables (for example, ages remain specific numerical values, and categorical variables retain their original categories). However, adding noise at the record level can significantly distort the data if it is excessive, or prove insufficient from a privacy point of view if it is limited. In this context, the main challenge involves obtaining an appropriate balance between privacy and utility.
The same applies if the focus is exclusively on publishing microdata (both via anonymization or sanitization using LDP or metric privacy). 

Thus, from the perspective of training machine or deep learning (ML/DL) models on such data, it is important to note that certain anonymization techniques are designed to prevent inference attacks, which may conflict with modeling objectives. For example, in the case of $\ell$-diversity, the goal is for individuals with the same QIs to have different values for the sensitive attribute, which can make it difficult for the model to learn consistent predictive relationships. In addition, the approach achieved with LDP-based approaches to microdata releasing under formal privacy guarantees allows the data to be reused directly in analytical workflows in which models have been trained with data under the same conditions, whereas if we deal with anonymized versions obtained using generalizations, such models could not be directly reused.

Furthermore, focusing on privacy-preserving machine learning (PPML), in horizontal federated learning (FL) scenarios \cite{el2022differential} where there are multiple clients with the same types of distributed data, the application of anonymization techniques requires ensuring consistency among the different participating nodes\cite{sainzpardo2024anjana}. Differences in generalization levels (for example, using exact ages in one node and age ranges in another or lat-lon intervals instead of exact points) can lead to incompatible models and degrade the performance of the overall model. These types of limitations are partially mitigated by using LDP to generate homogeneous noised versions of the data instead of applying generalizations. In addition, DP could be added at the client level also to disturb local model updates \cite{hu2024overview}. 

It is worth noting that these approaches are not incompatible with the application of global differential privacy. For example, in statistical analysis, mechanisms such as Laplace or Gaussian can be applied to protect aggregated queries, while techniques such as k-ary Randomized Response allow for the private estimation of distributions. Similarly, during the training of deep learning models, differential privacy can be incorporated using algorithms such as DP-SGD, together with specific versions of different optimizers with DP (e.g. Adam) implemented in libraries like \texttt{TensorFlow Privacy} \cite{abadi2016deep}. In the context of federated learning, differential privacy can also be applied during the aggregation phase of local models from the server side to prevent the extraction of sensitive information by potentially malicious participants.

Overall, the implementation of \texttt{trasgoDP} enables its integration into a complete data science workflow, facilitating the generation of noised data with local DP and metric privacy guarantees, and even its combination with global differential privacy mechanisms during model analysis and training. In this way, such integration with these other solutions allows to protect not only the privacy of the input data but also that of the resulting model, both in inference scenarios and during production deployment. Whether to use this tool or anonymization ones depends on the specific use case, the application to be developed with that data (model development, publication, federated training, etc), the number of records, their distribution, the desired level of privacy, and the types of attacks that need to be prevented.

\subsection*{Limitations of the study}

Certain limitations of the current version of \texttt{trasgoDP} (v2.0.3) should be acknowledged. First, regarding local differential privacy for tabular attributes, the library currently implements a limited set of mechanisms (two for numerical attributes, three for categorical ones and one for location-based data), and does not yet support other data types available often in tabular dataset, such as free-text attributes.

Second, the main limitation of applying LDP for microdata releasing remains its application to multiple columns of the same dataset: due to the sequential composition theorem, the overall privacy guarantee degrades according to the sum of the individual budgets. Since noise is added independently at the record level (local model), the reduction in utility loss observed as $\epsilon$ increases must be further explored. Thus, a more rigorous analysis of the privacy budget 
composition is left for future work.

Third, in this same line, the correlation loss metric proposed in this work and implemented in \texttt{trasgoDP}, has only been validated empirically on two datasets, and its behavior on datasets with different sizes, dimensionalities, or feature  correlation structures remains to be studied.

Finally, as with any LDP-based approach, results are sensitive to the choice of the privacy budget $\epsilon$ (and $\delta$ for the Gaussian mechanism). No automated procedure is currently provided within \texttt{trasgoDP} to select these parameters based on a target utility level, and this limitation remains as future work.

\section*{METHODS}

\paragraph{Methodology and development conventions}
The development of the \texttt{trasgoDP} library has been conducted openly on GitHub. Regarding the development methodology, the project adopts the \textit{Conventional Commits} specification, in which each commit is categorized using standardized prefixes such as \texttt{feat:}, \texttt{fix:}, \texttt{docs:}, \texttt{test:}, and \texttt{chore:} among others. As discussed below, this convention enables the automatic generation of changelogs and facilitates semantic versioning by determining the appropriate version increment (major, minor, or patch).

The project follows \textit{Semantic Versioning (SemVer)}, and the version referenced in this paper is v2.0.3. For this version, the package is classified under \textit{Development Status :: 5 - Production/Stable}, targeting developers, educators, and scientific researchers, and it is distributed under the Apache 2.0 license.

\paragraph{Dependencies management and installation}
The software development of \texttt{trasgoDP} relies on \texttt{Poetry}~\cite{poetry} for  dependency management and packaging, with all configuration centralized in the \texttt{pyproject.toml} file. Dependencies are organized into separate groups according to their purpose: development environment (\texttt{tox}), unit testing (\texttt{pytest}\cite{pytest} and \texttt{pytest-cov}), code linting and style enforcement (\texttt{flake8}\cite{flake8}), code formatting (\texttt{black}\cite{black}), static security analysis (\texttt{bandit}\cite{bandit}), static type checking (\texttt{mypy}), undeclared dependency detection (\texttt{pip-check-reqs}), and package validation prior to publication (\texttt{twine}\cite{twine}). This modular structure ensures that only the dependencies strictly required for each task are installed, preventing unnecessary installation in production environments. Finally, the project uses \texttt{tox} to coordinate all testing and quality assurance environments. Testing covers five versions of Python (from 3.10 to 3.14), ensuring wide compatibility. 

\paragraph{CI/CD and code coverage}
Continuous integration and delivery are managed through three dedicated GitHub Actions workflows. The main pipeline (\texttt{cicd.yml}) runs all the tests on every push or pull request. A separate workflow (\texttt{pypi.yml}) handles automated publication of the package in PyPI upon the creation of a new release, while \texttt{.codecov.yml} reports code coverage metrics to
Codecov~\cite{codecov}. Project documentation is automatically built using a dedicated pipeline and hosted via Read the Docs.

\paragraph{Unit testing}
As mentioned previously, unit testing has been conducted using the \textit{adult dataset}. The \texttt{unittest} framework is employed to validate both edge cases and the correct behavior of the implemented functions concerning types of both inputs and outputs. The current test suite achieves over 97$\%$ code coverage in version v2.0.3.

\paragraph{Release automation}  
Although the software can be installed directly via \texttt{pip} + \texttt{git} and referring to the original repository, the release process has been fully automated through PyPI distribution. The package can therefore be installed using the command \codehighlight{pip install trasgodp}. This automation is enabled by the GitHub Actions workflow \texttt{pypi.yml}, together with the CI/CD pipeline that manages package validation and deployment via \texttt{twine}.

\paragraph{Documentation} 
The documentation of \texttt{trasgoDP} is published on Read the Docs and built automatically via the aforementioned pipeline. It is generated using \texttt{Sphinx}\cite{brandl2021sphinx}, with \texttt{furo} as the documentation theme. The build process takes advantage of \texttt{autodoc}, extracting information from docstrings to describe function inputs, outputs, behavior, and expected types.

\paragraph{Requirements} 
Finally, with respect to version v2.0.3, beyond the dependencies required for testing, documentation, and code quality, the core runtime dependencies remain minimal, including the following: \texttt{numpy} (v2.0.2), \texttt{pandas} (v2.3.3), \texttt{scipy} (v1.15.3), \texttt{typing\_extensions}\cite{typing_extensions} (v4.16.0), \texttt{folium} (v0.20.0). This lightweight dependency design enhances compatibility with a wide range of Python ecosystems and ensures support across Python versions 3.10 through 3.14.

\section*{RESOURCE AVAILABILITY}

Requests for further information and resources should be directed to and will be fulfilled by the lead contact, Judith Sáinz-Pardo Díaz (sainzpardo@ifca.es).

\subsection*{Materials availability}

References to the code generated from this study are given in the following section. Only synthetic versions of openly available datasets have been generated during this work.

\subsection*{Data and code availability}

The raw data used to conduct this work and to perform the unit tests are available in \href{https://github.com/IFCA-Advanced-Computing/trasgoDP/tree/main/examples}{the examples folder of the library's repository}.

The most relevant links concerning the availability of the source code, documentation and installation of the \texttt{trasgoDP} library are listed below:
\begin{itemize}
    \item The code for the \texttt{trasgoDP} library is openly available in GitHub:
    \href{https://github.com/IFCA-Advanced-Computing/trasgodp}{https://github.com/IFCA-Advanced-Computing/trasgodp}.
    \item The documentation can be found in ReadTheDocs: \href{https://trasgodp.readthedocs.org}{https://trasgodp.readthedocs.org}.
    \item The installation can be managed using PyPI: \href{https://pypi.org/project/trasgoDP/}{https://pypi.org/project/trasgoDP/}.
    \item In order to ensure code reproducibility and availability, a Zenodo DOI has been created. This DOI stands for all the tagged versions, and always resolves to the latest one: \\
    \href{https://zenodo.org/records/18754857}{https://zenodo.org/records/18754857}.
\end{itemize}

Any additional information required concerning the code reported in this paper is available from the lead contact upon request. 

\section*{ACKNOWLEDGMENTS}

The authors would like to thank the funding through the EOSC SIESTA project ``Secure Interactive Environments for Sensitive daTa Analytics'', funded by the European Union (Horizon Europe) under grant agreement number 101131957, and the support from the EOSC ARENA project ``AI Research Enhancement through Networked Agents'', funded by the European Union (Horizon Europe) under grant agreement number 101292597.

\section*{AUTHOR CONTRIBUTIONS}

J.S-P.D. defined and conceived this work, performed the formal analysis, methodology, validation and software development. A.L.G. contributed to the definition and software development of this work and was responsible of supervision, project administration and funding acquisition. Both authors contributed to writing and reviewing the manuscript.

\section*{DECLARATION OF INTERESTS}

The authors declare that they have no known competing financial interests or personal relationships that could have appeared to influence the work reported in this paper.

\bibliographystyle{cas-model2-names}

\bibliography{arxiv/main_arxiv}

\begin{thebibliography}{36}
\expandafter\ifx\csname natexlab\endcsname\relax\def\natexlab#1{#1}\fi
\providecommand{\url}[1]{\texttt{#1}}
\providecommand{\href}[2]{#2}
\providecommand{\path}[1]{#1}
\providecommand{\DOIprefix}{doi:}
\providecommand{\ArXivprefix}{arXiv:}
\providecommand{\URLprefix}{URL: }
\providecommand{\Pubmedprefix}{pmid:}
\providecommand{\doi}[1]{\href{http://dx.doi.org/#1}{\path{#1}}}
\providecommand{\Pubmed}[1]{\href{pmid:#1}{\path{#1}}}
\providecommand{\bibinfo}[2]{#2}
\ifx\xfnm\relax \def\xfnm[#1]{\unskip,\space#1}\fi
\bibitem[{Abadi et~al.(2016)Abadi, Chu, Goodfellow, McMahan, Mironov, Talwar and Zhang}]{abadi2016deep}
\bibinfo{author}{Abadi, M.}, \bibinfo{author}{Chu, A.}, \bibinfo{author}{Goodfellow, I.}, \bibinfo{author}{McMahan, H.B.}, \bibinfo{author}{Mironov, I.}, \bibinfo{author}{Talwar, K.}, \bibinfo{author}{Zhang, L.}, \bibinfo{year}{2016}.
\newblock \bibinfo{title}{Deep learning with differential privacy}, in: \bibinfo{booktitle}{Proceedings of the 2016 ACM SIGSAC conference on computer and communications security}, pp. \bibinfo{pages}{308--318}.
\bibitem[{Andr\'{e}s et~al.(2013)Andr\'{e}s, Bordenabe, Chatzikokolakis and Palamidessi}]{andres2013geo}
\bibinfo{author}{Andr\'{e}s, M.E.}, \bibinfo{author}{Bordenabe, N.E.}, \bibinfo{author}{Chatzikokolakis, K.}, \bibinfo{author}{Palamidessi, C.}, \bibinfo{year}{2013}.
\newblock \bibinfo{title}{Geo-indistinguishability: differential privacy for location-based systems}, in: \bibinfo{booktitle}{Proceedings of the 2013 ACM SIGSAC Conference on Computer \& Communications Security}, \bibinfo{publisher}{Association for Computing Machinery}, \bibinfo{address}{New York, NY, USA}. p. \bibinfo{pages}{901–914}.
\newblock \URLprefix \url{https://doi.org/10.1145/2508859.2516735}, \DOIprefix\doi{10.1145/2508859.2516735}.
\bibitem[{Arcolezi et~al.(2022)Arcolezi, Couchot, Gambs, Palamidessi and Zolfaghari}]{arcolezi2022multi}
\bibinfo{author}{Arcolezi, H.H.}, \bibinfo{author}{Couchot, J.F.}, \bibinfo{author}{Gambs, S.}, \bibinfo{author}{Palamidessi, C.}, \bibinfo{author}{Zolfaghari, M.}, \bibinfo{year}{2022}.
\newblock \bibinfo{title}{Multi-freq-ldpy: multiple frequency estimation under local differential privacy in python}, in: \bibinfo{booktitle}{European Symposium on Research in Computer Security}, \bibinfo{organization}{Springer}. pp. \bibinfo{pages}{770--775}.
\bibitem[{Becker and Kohavi(1996)}]{adult_dataset}
\bibinfo{author}{Becker, B.}, \bibinfo{author}{Kohavi, R.}, \bibinfo{year}{1996}.
\newblock \bibinfo{title}{{Adult}}.
\newblock \bibinfo{howpublished}{UCI Machine Learning Repository}.
\newblock \bibinfo{note}{{DOI}: https://doi.org/10.24432/C5XW20}.
\bibitem[{Bertram et~al.(2023)Bertram, Sundin, Roche, S{\'a}nchez-T{\'o}jar, Thor{\'e} and Brodin}]{bertram2023open}
\bibinfo{author}{Bertram, M.G.}, \bibinfo{author}{Sundin, J.}, \bibinfo{author}{Roche, D.G.}, \bibinfo{author}{S{\'a}nchez-T{\'o}jar, A.}, \bibinfo{author}{Thor{\'e}, E.S.}, \bibinfo{author}{Brodin, T.}, \bibinfo{year}{2023}.
\newblock \bibinfo{title}{Open science}.
\newblock \bibinfo{journal}{Current biology} \bibinfo{volume}{33}, \bibinfo{pages}{R792--R797}.
\bibitem[{Biswas and Palamidessi(2024)}]{biswas2024privic}
\bibinfo{author}{Biswas, S.}, \bibinfo{author}{Palamidessi, C.}, \bibinfo{year}{2024}.
\newblock \bibinfo{title}{Privic: A privacy-preserving method for incremental collection of location data}.
\newblock \bibinfo{journal}{Proceedings on Privacy Enhancing Technologies} .
\bibitem[{Chatzikokolakis et~al.(2013)Chatzikokolakis, Andr{\'e}s, Bordenabe and Palamidessi}]{chatzikokolakis2013broadening}
\bibinfo{author}{Chatzikokolakis, K.}, \bibinfo{author}{Andr{\'e}s, M.E.}, \bibinfo{author}{Bordenabe, N.E.}, \bibinfo{author}{Palamidessi, C.}, \bibinfo{year}{2013}.
\newblock \bibinfo{title}{Broadening the scope of differential privacy using metrics}, in: \bibinfo{booktitle}{international symposium on privacy enhancing technologies symposium}, \bibinfo{organization}{Springer}. pp. \bibinfo{pages}{82--102}.
\bibitem[{Ciriani et~al.(2007)Ciriani, De~Capitani~di Vimercati, Foresti and Samarati}]{ciriani2007microdata}
\bibinfo{author}{Ciriani, V.}, \bibinfo{author}{De~Capitani~di Vimercati, S.}, \bibinfo{author}{Foresti, S.}, \bibinfo{author}{Samarati, P.}, \bibinfo{year}{2007}.
\newblock \bibinfo{title}{Microdata protection}, in: \bibinfo{booktitle}{Secure data management in decentralized systems}. \bibinfo{publisher}{Springer}, pp. \bibinfo{pages}{291--321}.
\bibitem[{{City of New York}(2026)}]{nyc_taxis}
\bibinfo{author}{{City of New York}}, \bibinfo{year}{2026}.
\newblock \bibinfo{title}{{TLC Trip Record Data}}.
\newblock \bibinfo{howpublished}{\url{https://www.nyc.gov/site/tlc/about/tlc-trip-record-data.page }}.
\newblock \bibinfo{note}{[Accessed 17-07-2026]}.
\bibitem[{Cormode et~al.(2021)Cormode, Maddock and Maple}]{cormode2021frequency}
\bibinfo{author}{Cormode, G.}, \bibinfo{author}{Maddock, S.}, \bibinfo{author}{Maple, C.}, \bibinfo{year}{2021}.
\newblock \bibinfo{title}{Frequency estimation under local differential privacy [experiments, analysis and benchmarks]}.
\newblock \bibinfo{journal}{arXiv preprint arXiv:2103.16640} .
\bibitem[{Domingo-Ferrer et~al.(2016)Domingo-Ferrer, S{\'a}nchez and Soria-Comas}]{domingo2016database}
\bibinfo{author}{Domingo-Ferrer, J.}, \bibinfo{author}{S{\'a}nchez, D.}, \bibinfo{author}{Soria-Comas, J.}, \bibinfo{year}{2016}.
\newblock \bibinfo{title}{Database anonymization: privacy models, data utility, and microaggregation-based inter-model connections}.
\newblock \bibinfo{publisher}{Morgan \& Claypool Publishers}.
\bibitem[{Domingo-Ferrer et~al.(2022)Domingo-Ferrer, S{\'a}nchez and Soria-Comas}]{domingo2022anonymization}
\bibinfo{author}{Domingo-Ferrer, J.}, \bibinfo{author}{S{\'a}nchez, D.}, \bibinfo{author}{Soria-Comas, J.}, \bibinfo{year}{2022}.
\newblock \bibinfo{title}{Anonymization methods for microdata}, in: \bibinfo{booktitle}{Database Anonymization: Privacy Models, Data Utility, and Microaggregation-based Inter-model Connections}. \bibinfo{publisher}{Springer}, pp. \bibinfo{pages}{15--23}.
\bibitem[{Douriez et~al.(2016)Douriez, Doraiswamy, Freire and Silva}]{7796899}
\bibinfo{author}{Douriez, M.}, \bibinfo{author}{Doraiswamy, H.}, \bibinfo{author}{Freire, J.}, \bibinfo{author}{Silva, C.T.}, \bibinfo{year}{2016}.
\newblock \bibinfo{title}{Anonymizing nyc taxi data: Does it matter?}, in: \bibinfo{booktitle}{2016 IEEE International Conference on Data Science and Advanced Analytics (DSAA)}, pp. \bibinfo{pages}{140--148}.
\newblock \DOIprefix\doi{10.1109/DSAA.2016.21}.
\bibitem[{Dwork and Roth(2014)}]{dwork2014algorithmic}
\bibinfo{author}{Dwork, C.}, \bibinfo{author}{Roth, A.}, \bibinfo{year}{2014}.
\newblock \bibinfo{title}{The algorithmic foundations of differential privacy}.
\newblock \bibinfo{journal}{Foundations and trends in theoretical computer science} \bibinfo{volume}{9}, \bibinfo{pages}{211--487}.
\bibitem[{El~Ouadrhiri and Abdelhadi(2022)}]{el2022differential}
\bibinfo{author}{El~Ouadrhiri, A.}, \bibinfo{author}{Abdelhadi, A.}, \bibinfo{year}{2022}.
\newblock \bibinfo{title}{Differential privacy for deep and federated learning: A survey}.
\newblock \bibinfo{journal}{IEEE access} \bibinfo{volume}{10}, \bibinfo{pages}{22359--22380}.
\bibitem[{{Feroze, Zahid}(2026)}]{global_cancer_patients_2015_2024}
\bibinfo{author}{{Feroze, Zahid}}, \bibinfo{year}{2026}.
\newblock \bibinfo{title}{{Dataset: Global Cancer Patients 2015-2024}}.
\newblock \bibinfo{howpublished}{\url{https://www.kaggle.com/datasets/zahidmughal2343/global-cancer-patients-2015-2024}}.
\newblock \bibinfo{note}{[Accessed 09-02-2026]}.
\bibitem[{{Flake8 developers}(2026)}]{flake8}
\bibinfo{author}{{Flake8 developers}}, \bibinfo{year}{2026}.
\newblock \bibinfo{title}{Github repository: flake8}.
\newblock \bibinfo{howpublished}{\url{https://github.com/pycqa/flake8}}.
\newblock \bibinfo{note}{[Accessed 26-03-2026]}.
\bibitem[{Holohan et~al.(2019)Holohan, Braghin, Mac~Aonghusa and Levacher}]{diffprivlib}
\bibinfo{author}{Holohan, N.}, \bibinfo{author}{Braghin, S.}, \bibinfo{author}{Mac~Aonghusa, P.}, \bibinfo{author}{Levacher, K.}, \bibinfo{year}{2019}.
\newblock \bibinfo{title}{Diffprivlib: the {IBM} differential privacy library}.
\newblock \bibinfo{journal}{ArXiv e-prints} \bibinfo{volume}{1907.02444 [cs.CR]}.
\bibitem[{Hu et~al.(2024)Hu, Gong, Zhang, Seng, Xia and Jiang}]{hu2024overview}
\bibinfo{author}{Hu, K.}, \bibinfo{author}{Gong, S.}, \bibinfo{author}{Zhang, Q.}, \bibinfo{author}{Seng, C.}, \bibinfo{author}{Xia, M.}, \bibinfo{author}{Jiang, S.}, \bibinfo{year}{2024}.
\newblock \bibinfo{title}{An overview of implementing security and privacy in federated learning}.
\newblock \bibinfo{journal}{Artificial intelligence review} \bibinfo{volume}{57}, \bibinfo{pages}{204}.
\bibitem[{Mironov(2017)}]{8049725}
\bibinfo{author}{Mironov, I.}, \bibinfo{year}{2017}.
\newblock \bibinfo{title}{Rényi differential privacy}, in: \bibinfo{booktitle}{2017 IEEE 30th Computer Security Foundations Symposium (CSF)}, pp. \bibinfo{pages}{263--275}.
\newblock \DOIprefix\doi{10.1109/CSF.2017.11}.
\bibitem[{{OpenMined}(2026)}]{pydp}
\bibinfo{author}{{OpenMined}}, \bibinfo{year}{2026}.
\newblock \bibinfo{title}{Github repository: pydp}.
\newblock \bibinfo{howpublished}{\url{https://github.com/OpenMined/PyDP}}.
\newblock \bibinfo{note}{[Accessed 07-04-2026]}.
\bibitem[{Prasser et~al.(2020)Prasser, Eicher, Spengler, Bild and Kuhn}]{Prasser_Flexible_data_anonymization_2020}
\bibinfo{author}{Prasser, F.}, \bibinfo{author}{Eicher, J.}, \bibinfo{author}{Spengler, H.}, \bibinfo{author}{Bild, R.}, \bibinfo{author}{Kuhn, K.A.}, \bibinfo{year}{2020}.
\newblock \bibinfo{title}{{Flexible data anonymization using ARX - Current status and challenges ahead}}.
\newblock \bibinfo{journal}{Software: Practice and Experience} \bibinfo{volume}{50}, \bibinfo{pages}{1277--1304}.
\newblock \DOIprefix\doi{10.1002/spe.2812}.
\bibitem[{{Pytest-dev}(2026)}]{pytest}
\bibinfo{author}{{Pytest-dev}}, \bibinfo{year}{2026}.
\newblock \bibinfo{title}{Github repository: pytest}.
\newblock \bibinfo{howpublished}{\url{https://github.com/pytest-dev/pytest}}.
\newblock \bibinfo{note}{[Accessed 26-03-2026]}.
\bibitem[{{Python}(2026)}]{typing_extensions}
\bibinfo{author}{{Python}}, \bibinfo{year}{2026}.
\newblock \bibinfo{title}{Github repository: typing\_extensions}.
\newblock \bibinfo{howpublished}{\url{https://github.com/python/typing_extensions}}.
\newblock \bibinfo{note}{[Accessed 26-03-2026]}.
\bibitem[{{Python Code Quality Authority}(2026)}]{bandit}
\bibinfo{author}{{Python Code Quality Authority}}, \bibinfo{year}{2026}.
\newblock \bibinfo{title}{Github repository: bandit}.
\newblock \bibinfo{howpublished}{\url{https://github.com/PyCQA/bandit}}.
\newblock \bibinfo{note}{[Accessed 26-03-2026]}.
\bibitem[{{Python Packaging Authority}(2026)}]{twine}
\bibinfo{author}{{Python Packaging Authority}}, \bibinfo{year}{2026}.
\newblock \bibinfo{title}{Github repository: twine}.
\newblock \bibinfo{howpublished}{\url{https://github.com/pypa/twine}}.
\newblock \bibinfo{note}{[Accessed 26-03-2026]}.
\bibitem[{{Python-poetry}(2026)}]{poetry}
\bibinfo{author}{{Python-poetry}}, \bibinfo{year}{2026}.
\newblock \bibinfo{title}{Github repository: poetry}.
\newblock \bibinfo{howpublished}{\url{https://github.com/python-poetry/poetry}}.
\newblock \bibinfo{note}{[Accessed 26-03-2026]}.
\bibitem[{{Python Software Foundation}(2026)}]{black}
\bibinfo{author}{{Python Software Foundation}}, \bibinfo{year}{2026}.
\newblock \bibinfo{title}{Github repository: black}.
\newblock \bibinfo{howpublished}{\url{https://github.com/psf/black}}.
\newblock \bibinfo{note}{[Accessed 26-03-2026]}.
\bibitem[{Ramachandran et~al.(2021)Ramachandran, Bugbee and Murphy}]{ramachandran2021open}
\bibinfo{author}{Ramachandran, R.}, \bibinfo{author}{Bugbee, K.}, \bibinfo{author}{Murphy, K.}, \bibinfo{year}{2021}.
\newblock \bibinfo{title}{From open data to open science}.
\newblock \bibinfo{journal}{Earth and Space Science} \bibinfo{volume}{8}, \bibinfo{pages}{e2020EA001562}.
\bibitem[{Sainz-Pardo~Diaz and Lopez~Garcia(2022)}]{sainzpardo2022pycanon}
\bibinfo{author}{Sainz-Pardo~Diaz, J.}, \bibinfo{author}{Lopez~Garcia, A.}, \bibinfo{year}{2022}.
\newblock \bibinfo{title}{A python library to check the level of anonymity of a dataset}.
\newblock \bibinfo{journal}{Scientific Data} \bibinfo{volume}{9}, \bibinfo{pages}{785}.
\bibitem[{S{\'a}inz-Pardo~D{\'\i}az and L{\'o}pez~Garc{\'\i}a(2024)}]{sainzpardo2024anjana}
\bibinfo{author}{S{\'a}inz-Pardo~D{\'\i}az, J.}, \bibinfo{author}{L{\'o}pez~Garc{\'\i}a, {\'A}.}, \bibinfo{year}{2024}.
\newblock \bibinfo{title}{An open source python library for anonymizing sensitive data}.
\newblock \bibinfo{journal}{Scientific data} \bibinfo{volume}{11}, \bibinfo{pages}{1289}.
\bibitem[{{Sentry}(2026)}]{codecov}
\bibinfo{author}{{Sentry}}, \bibinfo{year}{2026}.
\newblock \bibinfo{title}{Codecov}.
\newblock \bibinfo{howpublished}{\url{https://about.codecov.io/}}.
\newblock \bibinfo{note}{[Accessed 26-03-2026]}.
\bibitem[{Shoemate et~al.()Shoemate, Vyrros, McCallum, Prasad, Durbin, Casacuberta~Puig, Cowan, Xu, Ratliff, Berrios, Whitworth, Eliot, Lebeda, Renard and McKay~Bowen}]{Shoemate_OpenDP_Library}
\bibinfo{author}{Shoemate, M.}, \bibinfo{author}{Vyrros, A.}, \bibinfo{author}{McCallum, C.}, \bibinfo{author}{Prasad, R.}, \bibinfo{author}{Durbin, P.}, \bibinfo{author}{Casacuberta~Puig, S.}, \bibinfo{author}{Cowan, E.}, \bibinfo{author}{Xu, V.}, \bibinfo{author}{Ratliff, Z.}, \bibinfo{author}{Berrios, N.}, \bibinfo{author}{Whitworth, A.}, \bibinfo{author}{Eliot, M.}, \bibinfo{author}{Lebeda, C.}, \bibinfo{author}{Renard, O.}, \bibinfo{author}{McKay~Bowen, C.}, .
\newblock \bibinfo{title}{{OpenDP Library}}.
\newblock \URLprefix \url{https://github.com/opendp/opendp}.
\bibitem[{{Sphinx developers}(2026)}]{brandl2021sphinx}
\bibinfo{author}{{Sphinx developers}}, \bibinfo{year}{2026}.
\newblock \bibinfo{title}{Sphinx documentation}.
\newblock \bibinfo{howpublished}{\url{https://www.sphinx-doc.org/}}.
\newblock \bibinfo{note}{[Accessed 26-03-2026]}.
\bibitem[{Xiong et~al.(2020)Xiong, Liu, Li, Cai and Niu}]{xiong2020comprehensive}
\bibinfo{author}{Xiong, X.}, \bibinfo{author}{Liu, S.}, \bibinfo{author}{Li, D.}, \bibinfo{author}{Cai, Z.}, \bibinfo{author}{Niu, X.}, \bibinfo{year}{2020}.
\newblock \bibinfo{title}{A comprehensive survey on local differential privacy}.
\newblock \bibinfo{journal}{Security and Communication Networks} \bibinfo{volume}{2020}, \bibinfo{pages}{8829523}.
\bibitem[{Zhang et~al.(2025)Zhang, Mishra and Arcolezi}]{10.1145/3719027.3760706}
\bibinfo{author}{Zhang, H.}, \bibinfo{author}{Mishra, A.K.}, \bibinfo{author}{Arcolezi, H.H.}, \bibinfo{year}{2025}.
\newblock \bibinfo{title}{Demo: Exploring utility and attackability trade-offs in local differential privacy}, in: \bibinfo{booktitle}{Proceedings of the 2025 ACM SIGSAC Conference on Computer and Communications Security}, \bibinfo{publisher}{Association for Computing Machinery}, \bibinfo{address}{New York, NY, USA}. p. \bibinfo{pages}{4728–4730}.
\newblock \URLprefix \url{https://doi.org/10.1145/3719027.3760706}, \DOIprefix\doi{10.1145/3719027.3760706}.

\end{thebibliography}

\end{document}